\documentclass[12pt]{article}

\usepackage{fullpage}
\usepackage{graphicx}
\usepackage{amsmath,amssymb,amsfonts,amsthm,appendix}

\def\be{\begin{equation}}
\def\ee{\end{equation}}
\def\bea{\begin{eqnarray}}
\def\eea{\end{eqnarray}}
\def\bma{\begin{mathletters}}
\def\ema{\end{mathletters}}

\def\0{\overline{0}}

\def\q0{\underline{0}}

\newcommand{\C}{{\mathbb{C}}}

\def\tr{\mbox{tr}}
\def\one{\leavevmode\hbox{\small1\normalsize\kern-.33em1}}

\def\bra#1{\langle#1|} \def\ket#1{|#1\rangle}
\def\braket#1#2{\langle#1|#2\rangle}

\newcommand\openone{\leavevmode\hbox{\small1\kern-3.8pt\normalsize1}}

\newtheorem{theo}{Theorem}
\newtheorem{defin}[theo]{Definition}

\newtheorem{crit}[theo]{Criterion}
\newtheorem{lemma}[theo]{Lemma}
\newtheorem{prop}[theo]{Proposition}
\newtheorem{cor}[theo]{Corollary}

\begin{document}

\title{A convergent hierarchy of semidefinite programs\\
characterizing the set of quantum correlations}
\author{Miguel Navascu\'es$^{1,2}$, Stefano Pironio$^2$, Antonio Ac\'\i
n$^{2,3}$\\
\small $^1$Institute for Mathematical Sciences, Imperial College
London, SW7 2PG, United Kingdom\\
 \small $^2$ICFO-Institut de
Ciencies Fotoniques, 08860 Castelldefels
(Barcelona), Spain\\
\small $^3$ICREA-Instituci\'o Catalana de Recerca i Estudis Avan\c
cats, 08010 Barcelona, Spain}
\date{\today}
\maketitle
%%%%%%%%%%%% Abstract %%%%%%%%%%%%%%%%%%%%%%%%%%%

\begin{abstract}
We are interested in the problem of characterizing the correlations
that arise when performing local measurements on separate quantum
systems. In a previous work [\emph{Phys. Rev. Lett.} \textbf{98},
010401 (2007)], we introduced an infinite hierarchy of conditions
necessarily satisfied by any set of quantum correlations. Each of
these conditions could be tested using semidefinite programming. We
present here new results concerning this hierarchy. We prove in
particular that it is complete, in the sense that any set of
correlations satisfying every condition in the hierarchy has a
quantum representation in terms of commuting measurements. Although
our tests are conceived to rule out non-quantum correlations, and
can in principle certify that a set of correlations is quantum only
in the asymptotic limit where all tests are satisfied, we show that
in some cases it is possible to conclude that a given set of
correlations is quantum after performing only a finite number of
tests. We provide a criterion to detect when such a situation
arises, and we explain how to reconstruct the quantum states and
measurement operators reproducing the given correlations. Finally,
we present several applications of our approach. We use it in
particular to bound the quantum violation of various Bell
inequalities.
\end{abstract}

\section{Introduction}\label{introduction}

The main goal of Quantum Information Science (QIS) is to
understand the possibilities and limitations of the quantum
formalism for information processing and communication. Research
in QIS is concerned on one hand with the design of new protocols
exploiting the transmission and manipulation of information
encoded in quantum states (see for instance \cite{NC}). On the
other hand, it seeks to identify the constraints on information
processing imposed by the quantum formalism. For instance, various
information tasks, such as unconditionally secure bit commitment,
have been shown to be impossible in a quantum world
\cite{bitcomm}.

A standard scenario in QIS, and which serves as a primitive for
more complex protocols, consists of two distant, non-communicating
parties, conventionally called Alice and Bob, who share a quantum
system in a joint state $\rho$. Each party makes a measurement on
his share of the state and obtains a classical outcome. On a
phenomenological level, we may describe the situation by saying
that the two parties have access to a black box (see Figure 1).
\begin{figure}
  \centering
 \includegraphics[angle=-90,width=9 cm]{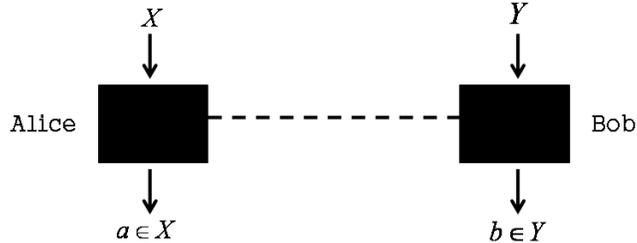}
  \caption{Local measurements on a system shared by two observers viewed as a black-box process. Alice chooses a measurement input $X$ and obtains a measurement output $a\in X$. Similarly, Bob chooses an input $Y$ and receives an output $b\in Y$. The behavior of the system is characterized by the joint probabilities $P(a,b)$.}
  \label{alicebob}
\end{figure}
When Alice inputs a measurement $X$ into the box, she gets as output
a measurement outcome $a\in X$; similarly, when Bob inputs a
measurement $Y$, he receives an output $b\in Y$. The behavior of the
box is completely characterized by the joint detection probabilities
$P(a,b)$. From now on, we simply call a \emph{behavior} the set
$P=\{P(a,b)\}$ of all such probabilities.

Though in the above scenario the parties are separated and perform
local measurements, their outcomes $a$ and $b$ may be
non-trivially correlated, in particular if the initial quantum
state $\rho$ is entangled. These correlated data can be exploited
for different tasks, such as communication
complexity~\cite{brassard} or key distribution~\cite{abgmps}. From
the perspective of QIS, it is thus meaningful to characterize
which outcome correlations can, or cannot, be produced by two
non-communicating quantum observers. The main problem with which
we are concerned in this paper is thus the following: given a
behavior $P$, do there exist a quantum state $\rho$ and local
measurements $X$ and $Y$ reproducing the outcome probabilities
described by $P$? Note that we do not impose here any constraints
on the dimension of the system shared by Alice and Bob, as we are
interested in the most general set of correlations that can be
obtained with quantum resources.

The special case of classical observers is relatively well
understood. The correlations obtained in this scenario coincide
with the ones that can be achieved with shared randomness, or
using another terminology, with those that are described by
local-hidden variable models~\cite{wernerwolf}. For a given number
of possible measurement inputs and outputs, the set of local
classical correlations forms a convex polytope whose vertices
correspond to all the possible deterministic assignments of
outputs to inputs. It thus follows that linear programming can be
used to decide if a given behavior is reproducible by two local
classical observers~\cite{zuklinprog,mprg}. The facets of the
classical polytope, which form the boundary of the classical
region, correspond to the well-known Bell
inequalities~\cite{Bell}.

Our understanding of the general case of quantum observers, with
which we are concerned here, is more rudimentary. The difficulty
lies in the fact that we do not have a practical characterization of
the set of quantum behaviors and that this set cannot be described
by a finite number of extreme points (see Figure 2)~\cite{tsir}.
\begin{figure}
  \centering
 \includegraphics[width=7 cm]{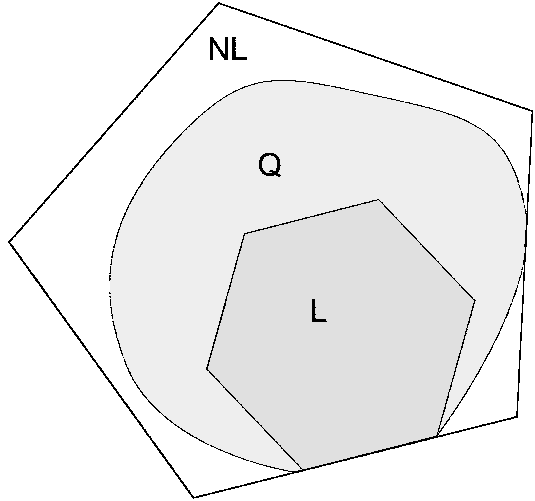}
  \caption{Schematic representations of the space of joint distributions $P(a,b)$ (for fixed and finite number of possible inputs and outputs). $L$ denotes the set of correlations that admit a local model; it
  is a polytope and membership in $L$ can be decided using linear programming
  \cite{zuklinprog,mprg}. $NL$ is the global set that contains all (in particular non-local) correlations; it is again a polytope. The region accessible to quantum mechanics is $Q$. The quantum set is not a polytope, i.e., it does not have a finite number of
  extreme points. As represented on the figure, $Q$ contains $L$
  and is a proper subset of $NL$. See \cite{tsir} or \cite{nosig} for more
  details.}
  \label{sets}
\end{figure}

Apart from the QIS motivation, the problem of characterizing the
set of quantum behaviors is also of relevance from a fundamental
perspective. Indeed, while Quantum Mechanics has been so far
confirmed by plenty of experiments, we cannot exclude that someday
it will be disproved. If some experimental data were inconsistent
with the quantum model that we have for the experiment, would that
however necessarily imply the breakdown of the whole quantum
formalism? How could one exclude that there is no other quantum
model explaining the observed data? The problem then is to
establish experimentally testable conditions that can rule out the
whole quantum structure, in a similar fashion as Bell's
inequalities do for locally causal model. In this context, one is
again confronted with the problem of characterizing the
constraints on correlations imposed by the quantum formalism.

One of the first researchers to study the characterization of
quantum correlations was Tsirelson in 1980 \cite{cir80}. Tsirelson
got several important results for the case of measurements with
binary outcomes \cite{tsir}; most notably he derived the maximal
quantum violation of the Clauser-Horne-Shimony-Holt (CHSH)
inequality \cite{chsh}. More recently, the problem has attracted
the interest of several researchers working in QIS \cite{list}.
Among the latest contributions, we point out the work of Wehner
\cite{wehner}, who showed that part of Tsirelson's findings could
be implemented using a relatively new numerical tool called
semidefinite programming (SDP). A short introduction to this
technique is given in Appendix A, more details can be found in
\cite{sdp}. Apart from Wehner's paper, there are several other
articles using SDP techniques to bound the set of quantum
correlations \cite{avis,doherty}. Most of these results deal with
the case of two-outcome measurements.

In a recent work \cite{anterior}, we introduced a hierarchy of SDP
tests to check if a given behavior admits a quantum representation.
Compared to previous constructions, our method is completely general
as it can be applied to any number of parties, measurements, and
outcomes, and is independent of the dimension of the quantum
systems. In this work, we explore further the approach introduced in
\cite{anterior}.

The basic idea behind our method is presented in
Section~\ref{basicidea}. Instead of directly searching for a
quantum state and measurement operators reproducing a given
behavior --- a computationally highly difficult task, if not
impossible if the dimension of the system is unbounded~--- we
consider instead a family of weaker conditions. Each of our
conditions amounts to verify the existence of a positive
semidefinite matrix whose structure depends on the general
algebraic properties satisfied by quantum states and measurement
operators. If one of our conditions is not satisfied, we can
immediately conclude that the given behavior is not quantum. In
Section~\ref{hierarchy}, we show that our family of tests can be
organized as an infinite hierarchy of increasingly stronger
conditions. We prove that in the asymptotic limit, our hierarchy
is complete in the sense that any behavior that satisfies all the
conditions in the hierarchy necessarily has a quantum
representation in terms of commuting measurements. We show further
that in some cases it is possible to conclude that a behavior is
quantum after a \emph{finite} number of steps only. We provide a
criterion to detect when such a situation arises, and we explain
how to reconstruct in this case explicit quantum states and
measurement operators reproducing the given behavior. Based on
these latter results, we then show how a slight modification of
our tests allow us to reduce the problem of deciding if a behavior
has a quantum representation with quantum systems of finite
dimension $d$ to a rank minimization problem. Unfortunately, and
contrary to SDP, there are no efficient algorithms to solve
rank-minimization problems. In Section~\ref{applications}, we
present several applications of our method. We use it in
particular to put upper bounds on the quantum violation of various
Bell inequalities. We conclude with a discussion and some open
questions in Section \ref{conclusion}.

\section{Definitions}
\label{definitions}

\subsection{Measurement scenario}
We consider measurement scenarios as illustrated in Figure 1. We
assume that outputs corresponding to different inputs are labeled
in a distinct way. Each output, say $a$ of Alice, is thus uniquely
associated to a single input $X(a)$. We denote by $A$ the set of
all outputs of Alice and by $B$ the set of all outputs of Bob. The
inputs of Alice may be viewed as disjoint subsets of $A$, and
those of Bob as disjoints subsets of $B$. A measurement scenario
is thus specified by a quadruple $(A,B,\mathcal{X},\mathcal{Y})$,
where $\mathcal{X}$ and $\mathcal{Y}$ are partitions of $A$ and
$B$, respectively.

The measurement scenarios that we consider in this paper always
involve a \emph{finite} number of inputs and outputs, i.e., $A$
and $B$ are finite sets. A behavior $P$ thus consists of a finite
set of $|A|\times |B|$ joint probabilities: $P=\{P(a,b)\,:\, a \in
A, b\in B\}$. For instance, in the case where Alice and Bob have
each a choice between $s$ different inputs that each yield one out
of $d$ outputs, a behavior consists of $s^2\times d^2$ joint
probabilities. Except otherwise mentioned, we assume in the
remaining of the paper that a measurement scenario
$(A,B,\mathcal{X},\mathcal{Y})$ and a behavior $P$ have been
specified. Our aim is to determine if $P$ represents a possible
quantum process.

\subsection{Quantum behaviors}
\label{defsets}
\begin{defin}
\label{defcomm} The behavior $P$ is a quantum
behavior if there exists a pure state $\ket{\psi}$ in a Hilbert
space $\mathcal{H}$, a set of measurement operators $\{E_a\,:\,
a\in A\}$ for Alice, and a set of measurement operators
$\{E_b\,:\, b\in B\}$ for Bob, such that for all $a\in A$ and
$b\in B$
\be\label{funda}
P(a,b)=\bra{\psi}E_a \,E_b\,\ket{\psi}\,,
\ee
with the measurement operators $E$ satisfying
\begin{enumerate}
\item $E_a^\dagger=E_a$ and $E_b^\dagger=E_b$
(hermiticity)
\item $E_a E_{\bar{a}}=\delta_{a\bar a} E_a$ if $X(a)=X(a')$ and $E_b E_{\bar{b}}=\delta_{b\bar b} E_b$ if $Y(b)=Y(b')$
(orthogonality)
\item $\sum_{a\in
X}E_a=\openone$ and $\sum_{b\in Y}E_b=\openone$ (completeness)
\item $[E_a,E_b]=0$ (commutativity)
\end{enumerate}
The set of all quantum behaviors will be denoted by $Q$.
\end{defin}
The first three properties are necessary to ensure that the
operators $E_a$ and $E_b$ are projectors and define proper
measurements. The fourth property simply expresses the fact that
Alice and Bob perform separated measurements on the global state
$\ket{\psi}$.

Note that more generally we could have defined a quantum behavior
in terms of a mixed state $\rho$ rather than a pure one and in
terms of general measurements, also known as Positive Operator
Valued Measures (POVM) \cite{NC}, rather than projective ones. But
remark also that in our definition we put no restrictions at all
on the dimension of the Hilbert space. Since any general
measurement on a given Hilbert space can be viewed as a projective
measurement on a larger Hilbert space, and any mixed state $\rho$
can be viewed as a subsystem of a larger system in a pure state
$\ket{\psi}$ \cite{NC}, the above definition turns out to be
completely general.

Property \emph{3} implies that the marginal probabilities
$P(a)=\sum_{b\in Y}P(a,b)$ and $P(b)=\sum_{a\in X}P(a,b)$ are
well-defined and independent of what is measured on the other side
(i.e., $P$ satisfies the no-signalling constraints). This also
implies that in the above definition there is some redundancy in the
specification of the operators $\{E_a \,:\,a\in
 A\}$ of Alice since any one of them can be written as the
identity minus the other ones. To simplify further the definition
above, select an output $a_X\in X$ for each input $X$ and
introduce the reduced output sets $\tilde X=\{a\,:\, a\in X, a\neq
a_X\}$ and $\tilde A=\bigcup_X \tilde X$. Introduce analogous sets
$\tilde Y$ and $\tilde B$ for Bob. The following definition is
then equivalent to Definition \ref{defcomm}.
\begin{defin}
\label{defcomm2}
The behavior $P$ is a quantum behavior if there
exists a pure (normalized) state $\ket{\psi}$ in a Hilbert space
$\mathcal{H}$, a set of measurement operators $\{E_a\,:\, a\in
\tilde A\}$ for Alice, and a set of measurement operators
$\{E_b\,:\, b\in \tilde B\}$ for Bob such that for all $a\in\tilde
A$ and $b\in \tilde B$
\begin{alignat}{2}
P(a)&=\bra{\psi}E_a\ket{\psi}\nonumber\\
P(b)&=\bra{\psi}E_b\ket{\psi}\nonumber\\
P(a,b)&=\bra{\psi}E_a E_b\ket{\psi}\label{funda2}
\end{alignat}
with the measurement operators satisfying
\begin{enumerate}
\item $E_a^\dagger=E_a$ and $E_b^\dagger=E_b$
(hermiticity)
\item $E_a E_{\bar{a}}=\delta_{a\bar a} E_a$ if $X(a)=X(a')$ and $E_b E_{\bar{b}}=\delta_{b\bar b} E_b$ if $Y(b)=Y(b')$
(orthogonality)
\item $[E_a,E_b]=0$ (commutativity)
\end{enumerate}
\end{defin}
It is clear that any behavior satisfying Definition \ref{defcomm}
also satisfies Definition \ref{defcomm2}. The converse statement
is also true. Indeed given sets of operators $\{E_a\,:\,
a\in\tilde A\}$ and $\{E_b\,:\, b\in\tilde B\}$ satisfying
Definition \ref{defcomm2}, define the missing operators $E_{a_X}$
and $E_{b_Y}$ through $E_{a_X}=\openone-\sum_{a\in\tilde X}E_a$
and $E_{b_Y}=\openone-\sum_{b\in\tilde Y}E_b$. It is then easy to
see that the now-complete sets $\{E_a\,:\, a\in A\}$ and
$\{E_b\,:\, b\in B\}$ satisfy Definition \ref{defcomm}.

Before concluding this subsection, note that when dealing with
finite dimensional Hilbert spaces, one tends to associate a tensor
product structure to separated measurements. This leads to another
set $Q'$ of quantum behaviors, possibly equivalent to $Q$, and
defined as follows.
\begin{defin}
\label{deftens}
The behavior $P$ belongs to the set of quantum behaviors $Q'$ if there exists a pure state $\ket{\psi}$ in a composite Hilbert
space $\mathcal{H_A}\otimes\mathcal{H_B}$,
a set of measurement operators $\{E_a\,:\,
a\in A\}$ for Alice, and a set of measurement operators
$\{E_b\,:\, b\in B\}$ for Bob, such that for all $a\in A$ and
$b\in B$
\be\label{fundatens}
P(a,b)=\bra{\psi}E_a \otimes E_b\,\ket{\psi}\,,
\ee
with the measurement operators $E$ satisfying
\begin{enumerate}
\item $E_a^\dagger=E_a$ and $E_b^\dagger=E_b$
(hermiticity)
\item $E_a E_{\bar{a}}=\delta_{a\bar a} E_a$ if $X(a)=X(a')$ and $E_b E_{\bar{b}}=\delta_{b\bar b} E_b$ if $Y(b)=Y(b')$
(orthogonality)
\item $\sum_{a\in
X}E_a=\openone_A$ and $\sum_{b\in Y}E_b=\openone_B$ (completeness)
\end{enumerate}
\end{defin}
Clearly, $Q'\subseteq Q$. However, it is an open question whether
these two sets are equal. In the special case of finite
dimensional Hilbert spaces, they turn out to be
identical~\cite{openproblem}. In this work, we adopt
Definition~\ref{defcomm}, or equivalently
Definition~\ref{defcomm2}, partly because it is much better
tailored to the structure of our construction. We will come back
to the commutation versus tensor product issue in section
\ref{tensor}.

\subsection{Sets of operators and sequences}\label{seqop}

In this subsection, we introduce a few other definitions that will
be needed later on.

Let $\mathcal{E}$ denote the set of projectors appearing in
Definition \ref{defcomm}, i.e., $\mathcal{E}=\{E_a\,:\, a\in
A\}\cup\{E_b\,:\, b\in B\}$, and $\tilde{\mathcal{E}}$ denote the
set of projectors of Definition \ref{defcomm2} plus the identity,
i.e., $\tilde{\mathcal{E}}=\openone\cup\{E_a\,:\, a\in \tilde
A\}\cup\{E_b\,:\, b\in \tilde B\}$.

Let $\mathcal{O}=\{O_1,\ldots,O_n\}$ be a set of $n$ operators,
where each $O_i$ is a linear combination of products of projectors
in $\tilde{\mathcal{E}}$. Thus $\mathcal{O}$ is a finite subset of
the algebra generated by $\tilde{\mathcal{E}}$. Note that we can
equally well define the set $\mathcal{O}$ in terms of
$\mathcal{E}$, since $\mathcal{E}$ and $\tilde{\mathcal{E}}$ are
equivalent up to linear combinations. Define
$\mathcal{F}(\mathcal{O})$ as the set of all independent
equalities of the form
\be\label{eqop} \sum_{ij} {(F_k)}_{ij} \bra{\psi}O^\dagger_i
O_j\ket{\psi}=g_k\left(P\right) \qquad k=1,\ldots,m \ee which are
satisfied by the operators $O_i$, where the coefficients $g_k(P)$
are linear functions of the probabilities $P(a,b)$: \be\label{dk}
g_k(P)=(g_k)_0+\sum_{a,b} {(g_k)}_{ab} P(a,b) \ee and where
$\ket{\psi}$ is the state appearing in Definition \ref{defcomm2}.
These equations are the ones that formally follow from the
definition of the $O_i$'s, the relation \eqref{funda2}, and
properties \emph{1-3} of Definition \ref{defcomm2}. Each set of
operators $\mathcal{O}$ define such a collection of equations. As
an example of equation of the form \eqref{eqop}, suppose that the
set $\mathcal{O}$ contains the operators $\{O_k\}_{k=1}^d=\{E_{b}
E_{a} S\,:\, {a}\in X\}$, where $S$ is some arbitrary operator in
the algebra generated by $\mathcal{E}$, and also contains the
operator $O_{d+1}=E_{b} S$. Then $\sum_{k=1}^d O_k^\dagger O_k
=\sum_{{a}\in X}(E_{b} E_{a} S)^\dagger E_{b} E_{a} S=\sum_{{a}\in
X} S^\dagger E_{a} E_{b} E_{b} E_{a} S=\sum_{{a}\in X} S^\dagger
E_{b} E_{a} E_{b} S=S^\dagger E_{b} E_{b} S=O_{d+1}^\dagger
O_{d+1}$, and thus \linebreak[1]$\sum_{k=1}^d
\linebreak[1]\bra{\psi} O_k^\dagger O_k \ket{\psi} - \bra{\psi}
O_{d+1}^\dagger O_{d+1} \ket{\psi}=0$.

Let a \emph{sequence} $S$ be a product of projectors in
$\tilde{\mathcal{E}}$. Examples of sequences are $E_a$ and
$E_aE_{a'}E_b$. Note that some sequences may correspond to the
null operator, for instance, $E_aE_a'=0$ if $a\neq a'$, and
$X(a)=X(a')$; in the following, when we speak of a sequence, we
always mean a non-null sequence. The \emph{length} $|S|$ of a
sequence is the minimum number of projectors needed to generate
it. For instance $|E_aE_bE_a|=|E_aE_aE_b|=|E_aE_b|=2$. By
convention, the length of the identity operator is $|\openone|=0$.
We define $\mathcal{S}_n$ to be the set of sequences of length
smaller than or equal to $n$ (excluding null sequences). Thus

%¡¡¡¡Atención!!!! He cambiado A(B) por \tilde{A}(\tilde{B})

\begin{align*}
\mathcal{S}_0&=\{\openone\}\\
\mathcal{S}_1&=\mathcal{S}_0\cup \{E_a\,:\, a\in
\tilde{A}\}\cup\{E_b\,:\, b\in
\tilde{B}\}\\
\mathcal{S}_2&=\mathcal{S}_0\cup\mathcal{S}_1\cup\{E_a E_{a'}\,:\,
a,a'\in \tilde{A}\}\cup\{E_bE_{b'}\,:\, b,b'\in
\tilde{B}\}\cup\{E_a E_b\,:\, a\in
\tilde{A}, b\in \tilde{B}\}\\
\mathcal{S}_3&=\ldots
\end{align*}
It is clear that $\mathcal{S}_1\subseteq
\mathcal{S}_2\subseteq\ldots$, and that any operator
$O_i\in\mathcal{O}$ can be written as a linear combination of
operators in  $\mathcal{S}_n$ for $n$ sufficiently large.

\section{Basic idea of our method}\label{basicidea}
The following proposition associates to each set of operators
$\mathcal{O}$ satisfying Eqs.~\eqref{eqop} a condition that
restricts the possible correlations that can arise between two
quantum observers.
\begin{prop}\label{propmet} Let $\mathcal{O}$ be a set of operators and $\mathcal{F}(\mathcal{O})$ the set of equations of the form \eqref{eqop} satisfied by operators in $\mathcal{O}$. Then, a necessary condition for a behavior $P$ to be
quantum is that there exists a complex hermitian $n\times n$
positive semidefinite matrix $\Gamma\succeq 0$ whose entries
$\Gamma_{ij}$ satisfy
\be\label{lingam}
\sum_{ij} {(F_k)}_{ij} \Gamma_{ij}=g_k(P)\\
\qquad k=1,\ldots,m
\ee
Moreover, if the coefficients $F_k$ and $g_k$ in \eqref{eqop} are
real, we can take $\Gamma$ to be real as well.
\end{prop}
\begin{proof}
If $P$ is quantum, there exist a state $\ket{\psi}$ and projectors
$E_a$ and $E_b$ as in Definition \ref{defcomm2}, and therefore
there also exist operators $O_i$ satisfying the relations
\eqref{eqop}. Then simply define the entries of the matrix
$\Gamma$ through \be \Gamma_{ij}=\bra{\psi}O^\dagger_i O_j
\ket{\psi} \ee Clearly, $\Gamma$ satisfies \eqref{lingam}.
Moreover, it is positive semidefinite since for all $v\in \C^n$
\be v^\dagger \Gamma v = \sum_{ij} v_i^* \Gamma_{ij} v_j=\sum_{ij}
v_i^* \bra{\psi}O^\dagger_i O_j \ket{\psi} v_j=\bra{\psi}V^\dagger
V \ket{\psi}\geq 0 \ee where $V=\sum_j v_j O_j$.

If the coefficients $F_k$ and $g_k$ in \eqref{eqop} are real,
redefine $\Gamma$ as $\left(\Gamma+\Gamma^*\right)/2$. Then
$\Gamma$ still is positive semidefinite and satisfies
\eqref{lingam}.
\end{proof}

We will call a \emph{certificate} associated to $\mathcal{O}$ to any
$n\times n$ positive semidefinite matrix $\Gamma$ satisfying the
linear constraints \eqref{lingam}. As an illustration, we now give
two examples of application of Proposition~\ref{propmet}.

\subsubsection*{Example 1} Consider a measurement scenario
where Alice has a choice between two measurements, $X=1$ or $X=2$,
to perform on her subsystem, and where both measurements yield
binary outcomes with values $\pm X$. Likewise, Bob has a choice
between two measurements, $Y=3$ or $Y=4$, with outcomes $\pm Y$.

The single-party measurement averages $C_X=P({+}X)-P({-}X)$ and
$C_Y=P(+Y)-P(-Y)$ together with the two-party correlation functions
$C_{XY}=P(+X,+Y)+P(-X,-Y)\linebreak[1]-P(+X,-Y)-P(-X,+Y)$ fully
determine the response of the joint system of Alice and Bob. The
observed data are thus characterized by the eight numbers
$\{C_1,C_2,C_3,C_4,\linebreak[1]C_{12},C_{13},C_{23},C_{24}\}$ which
are equivalent to the knowledge of the entire set of probabilities
$P(\pm X, \pm Y)$.

\begin{crit}
If the data observed by Alice and Bob represent the response of a
quantum system, there exists a real symmetric $5\times 5$ positive
semidefinite matrix $\Gamma\succeq 0$ of the form
\be\Gamma=\begin{pmatrix}
1 & C_1 & C_2 & C_3 & C_4\\
& 1& u& C_{13} & C_{14}\\
& & 1 & C_{23} & C_{24}\\
& & & 1 & v \\
& & & & 1
\end{pmatrix}
\label{gammae1}
\ee
where $u$ and $v$ are arbitrary entries. (We have only given the
upper triangular part of $\Gamma$ since it is symmetric.)
\end{crit}
\begin{proof}
If the data observed by Alice and Bob represent the response of a
quantum system, there exist a state $\ket{\psi}$, two projectors
$E_{\pm X}$ associated to each of the two measurements $X=1,2$ of
Alice and two projectors $E_{\pm Y}$ associated to each of the two
measurements $Y=3,4$ of Bob. Let
$\mathcal{O}=\{\sigma_0,\sigma_1,\ldots,\sigma_4\}$ where
$\sigma_0=\openone$ is the identity operator and
$\sigma_i=E_{+i}-E_{-i}$ $(i=1,\ldots,4)$. It is easily verified
from Eqs.~\eqref{funda} and properties \emph{1-4} that these
operators satisfy the equalities
\begin{alignat}{2}
\bra{\psi}\sigma_i^\dagger \sigma_i \ket{\psi}&=1 &\quad&i=0,\ldots,4\\
\bra{\psi}\sigma_0^\dagger \sigma_i \ket{\psi}&=C_i &&i=1,\ldots,4\\
\bra{\psi}\sigma_i^\dagger \sigma_j \ket{\psi}&=C_{ij}
&&i=1,2;\;j=3,4
\end{alignat}
which are the counterparts of Eqs.~\eqref{eqop}. It immediately
follows that the associated $5\times 5$ matrix
$\Gamma_{ij}=\bra{\psi}\sigma^\dagger_i \sigma_j\ket{\psi}$ has
the form \eqref{gammae1}. It can be taken real if we further
redefine $\Gamma$ as $(\Gamma+\Gamma^*)/2$.
\end{proof}

%Atención! Cambio X=1,...,m, por X=1,...,s

\subsubsection*{Example 2} Consider the case where Alice and Bob
have a choice between $s$ different measurements that each yield
one out of $d$ possible outcomes. The $s$ measurements of Alice
are labeled $X=1,\ldots,s$ and her $m=s\times d$ possible outcomes
are labeled $a=1,\ldots,m$, where outcomes in the range
$1+(k-1)d,\ldots,kd$ belong to the measurement $X=k$. Analogously,
the $s$ measurements of Bob are labeled $Y=s+1,\ldots,2s$ and his
$m=s\times d$ outcomes are $b=m+1,\ldots,2m$, where again outcomes
in the range $1+(k-1)d,\ldots,kd$ belong to the measurement $Y=k$.

This measurement scenario is characterized by the $m^2$ joint
probabilities~$P(a,b)$.
\begin{crit}
If the set of $m^2$ probabilities $P(a,b)$ admits a quantum
representation, there exists a $2m\times 2m$ real symmetric
positive semidefinite matrix $\Gamma\succeq 0$ of the form
\be\label{exgamma} \Gamma=\begin{pmatrix}
Q &P\\
P^T&R
\end{pmatrix}
\ee
where the submatrix $P$ is the $m\times m$ table of probabilities
with entries $P_{ab}=P(a,b)$, and where the submatrices $Q$ and
$R$ satisfy
\begin{alignat}{2}\label{q}
Q_{aa'}&=\delta_{aa'}P(a)&\qquad&\mathrm{if}\; X(a)=X(a')\\
R_{bb'}&=\delta_{bb'}P(b)&&\mathrm{if}\; Y(b)=Y(b')
\end{alignat}
\end{crit}
\begin{proof}
If the measurement scenario is a quantum measurement scenario,
there exist a quantum state $\ket{\psi}$, $m$ projectors $E_a$ for
Alice, and $m$ projectors $E_b$ for Bob satisfying the properties
of Definition \ref{defcomm}. Consider the set
$\mathcal{O}=\mathcal{E}=\{E_1,\ldots,E_m,\linebreak[1]E_{m+1},\ldots
E_{2m}\}$ consisting of the $m$ operators of Alice and the $m$
ones of Bob. They satisfy the equalities
\begin{alignat}{2}
\bra{\psi}E_a E_b \ket{\psi}&=P(a,b) &\nonumber\\
\bra{\psi} E_a E_a'\ket{\psi}&=\delta_{aa'}P(a) &\qquad& \text{if } X(a)=X(a')\nonumber\\
\bra{\psi} E_b E_b'\ket{\psi}&=\delta_{bb'}P(b) &\qquad& \text{if
}
 Y(b)=Y(b')
\end{alignat}
as implied by Eqs.~\eqref{funda} and property \emph{2}. It
immediately follows that the certificate $\Gamma$ associated to
$\mathcal{O}$ has the form \eqref{exgamma}.
\end{proof}

Note that the matrix \eqref{exgamma} can be thought of as a table
of probabilities, where $\Gamma_{ij}$ is the probability to obtain
the two outcomes $i,j\in\{1,\ldots,2m\}$. The only entries of this
matrix which are not specified are the entries
$\Gamma_{aa'}=Q_{aa'}$ associated to different measurements of
Alice, $X(a)\neq X(a')$, and the entries $\Gamma_{bb'}=R_{bb'}$
associated to different measurements of Bob, $Y(b)\neq Y(b')$.
This is coherent with our interpretation of $\Gamma$ since in a
quantum scenario these entries correspond to non-commuting
measurements performed on the same subsystem and are thus not
jointly observable. Nonetheless, if the correlations $P(a,b)$ have
a quantum origin it is possible to assign a numerical value to
these undetermined entries, namely $\bra{\psi}E_a
E_{a'}\ket{\psi}$ and $\bra{\psi}E_b E_{b'}\ket{\psi}$\footnote{Or
the real part of these expressions, if we take $\Gamma$ real.},
such as the overall matrix \eqref{exgamma} is positive
semidefinite.

\subsection{Testing the existence of a certificate with SDP}
Checking the existence of a certificate $\Gamma$, such as the ones
given in Examples 1 and 2, can be cast as a semidefinite program.
Indeed it amounts to solve the following problem
\begin{alignat}{2}
\mbox{maximize}\quad & \lambda\nonumber\\
\mbox{subject to}\quad &\mbox{tr}\left(F_k^T\,\Gamma\right)=g_k(P) \quad k=1,...,m\label{existence}\\
& \Gamma-\lambda \openone\succeq 0\nonumber
\end{alignat}
which after some elementary manipulations can be put in the form
\eqref{dualb}. A positive solution
$\lambda\nolinebreak[4]\geq\nolinebreak[4] 0$ to the above problem
implies that there exists a positive semidefinite matrix
$\Gamma\succeq \lambda I\succeq 0$ compatible with the linear
constraints \eqref{lingam}. A strictly negative solution $\lambda<0$
means that any matrix $\Gamma$ compatible with \eqref{lingam} is
necessarily negative definite and thus that the given behavior $P$
does not represent the outcome of a quantum experiment.

As mentioned in the Appendix, there exist many available programs
to solve problems of the type \eqref{existence}. Such programs
solve these problems both in their primal and dual forms. The dual
of \eqref{existence} is
\begin{alignat}{2}
\mbox{minimize}\quad & \sum_k y_k g_k(P)\nonumber\\
\mbox{subject to}\quad &F(y)=\sum_k y_k F_k^T \succeq 0\label{existencedual}\\
& \sum _k y_k \mbox{tr}(F_k^T)=1\nonumber
\end{alignat}
If a program returns a negative solution for the primal for a
given behavior $P^*$, it also yields a dual feasible point $y$
such that $\sum_k y_k g_k(P^*)<0$. This dual feasible point
provides a proof that the given behavior $P^*$ is not quantum; it
can be interpreted as a \emph{quantum Bell inequality} violated by
$P^*$ in the sense that $\sum_k y_k g_k(P) \geq 0$ is a linear
inequality satisfied by all quantum probabilities. Indeed, the
coefficients $g_k(P)$ defined in \eqref{dk} depend linearly on the
probabilities $P(a,b)$, and thus the expression $\sum y_k g_k(P)$
is a linear expression in the probabilities $P(a,b)$. Moreover,
from the second line of \eqref{existencedual}, we deduce that for
all behaviors $P$ having a positive certificate $\Gamma\succeq 0$,
in particular, for all quantum behaviors, this linear expression is positive:
$\sum_k y_k g_k(P)=\sum_k y_k\mbox{tr}\left(F_k^T\,
\Gamma\right)=\mbox{tr}(F(y)\Gamma)\geq 0$ since $\Gamma\succeq
0$. The behavior $P^*$, however, violates this inequality, $\sum_k
y_k g_k(P^*)<0$, which demonstrates that it does not belong to
$Q$.

\subsection{Equivalence between certificates}
Each set $\mathcal{O}$ of operators that we can write down yields
a different condition satisfied by quantum theory. However, not
all conditions built in this way are independent, as the following
lemma shows.
\begin{lemma}\label{lemeq}
Let $\mathcal{O}$ and ${\mathcal{O}'}$ be two sets of operators such
that every operator in $\mathcal{O}'$ is a linear combination of
operators in $\mathcal{O}$. Then, the existence of a certificate
$\Gamma$ associated to $\mathcal{O}$ (for a given $P$) implies the
existence of a certificate $\Gamma'$ associated to $\mathcal{O}'$.
\end{lemma}
\begin{proof}
By hypothesis, every operator $O'_i\in \mathcal{O}'$ can be
written as $O'_i=\sum_k C_{ik}O_k$, where $O_k \in\mathcal{O}$.
Define then $\Gamma'_{ij}\equiv\sum_{kl}
C_{ki}^*\Gamma_{kl}C_{lj}$. It is clear that $\Gamma'$ satisfies
the equalities \eqref{lingam} associated to $\mathcal{O'}$, given
that $\Gamma$ satisfies the ones associated to $\mathcal{O}$. We
also have that $\Gamma'=C^\dagger\Gamma C\succeq 0$, and thus
$\Gamma'$ is a certificate associated to $\mathcal{O}'$.
\end{proof}

The criterion of Example 2 for $s=2$ and $d=2$, for instance, is
equivalent to the one of Example 1, because the set of eight
operators
$\{E_{+1},E_{-1},E_{+2},E_{-2},E_{+3},E_{-3},E_{+4},E_{-4}\}$ is
linearly equivalent to the set of five operators
$\{\sigma_0,\sigma_1,\sigma_2,\sigma_3,\sigma_4\}$.

In numerical implementations, we have of course always interest to
use a criterion based on a set $\mathcal{O}$ of linearly
independent operators so as to minimize the size of the matrices
involved. Note also that to check systematically all the
conditions that follow from our approach, it is sufficient to
check the ones associated with the sets $\mathcal{S}_n$ defined in
Subsection~\ref{seqop} since they generate by linear combinations
all other possible operators. This point is made more precise in
the next section.

\section{A Hierarchy of necessary conditions}\label{hierarchy}
Motivated by the above lemma, define a \emph{certificate of order
$n$}, denoted $\Gamma^n$, as a certificate associated to the set
of operators $\mathcal{S}_n$. A certificate of order $n$ is thus a
$|\mathcal{S}_n|\times |\mathcal{S}_n|$ matrix and to index its
row and columns we will use symbols that are in direct
correspondence with the elements of $\mathcal{S}_n$. Sequence
operators $S$, $E_a$, $E_aS$, and $\openone$ will be associated
with row or column indices, $s$, $a$, $as$, and $1$, respectively.
We define the length $|s|$ of an index $s$ to be the length $|S|$
of the corresponding sequence $S$. A certificate $\Gamma^n$ is
thus a matrix with entries $\{\Gamma^n_{s,t}\,:\,|s|,|t|\leq n\}$,
which according to the proof of Proposition~\ref{propmet} may be
interpreted as $\Gamma^n_{s,t}=\bra{\psi}S^\dagger T\ket{\psi}$ if
$P$ is a quantum behavior.

From Proposition~\ref{propmet} and the definition of the set
$\mathcal{S}_n$, we deduce that $\Gamma^n$ is a real positive
semidefinite matrix that satisfies the linear equalities
\be\label{certprob}
\Gamma^n_{1,1}=1\,,\quad\Gamma^n_{1,a}=P(a)\,,\quad\Gamma^n_{1,b}=P(b)\,,\quad\Gamma^n_{a,b}=P(a,b)
\ee for all $a\in\tilde A$ and $b\in \tilde B$, and
\begin{alignat}{1}
%\Gamma^n_{s,t}&=0 \quad \text{if } S^\dagger T=0\nonumber\\
\Gamma^n_{s,t}=\Gamma^n_{u,v} \quad \text{if } S^\dagger
T=U^\dagger V \qquad\left(\Gamma^n_{s,t}=0\quad  \text{if }
S^\dagger T=0\right)\label{ordern}
\end{alignat}
for all $|s|,|t|,|u|,|v|\leq n$. Here the relations $S^\dagger
T=U^\dagger V$ (or $S^\dagger T=0$) are the ones that follow from
properties \emph{1-3} of Definition~2. For instance,
$\Gamma^n_{ab,a}=\Gamma^n_{1,ab}$, and $\Gamma^n_{ab,a'}=0$ if
$X(a)=X(a')$.

As we mentioned earlier, $\mathcal{S}_1\subseteq
\mathcal{S}_2\subseteq\ldots\subseteq
\mathcal{S}_n\subseteq\ldots$, and thus the family of certificates
$\Gamma^1, \Gamma^2,\ldots,\Gamma^n\ldots,$ represents a hierarchy
of conditions satisfied by quantum probabilities, where each
condition in the hierarchy is stronger than the previous ones.
Moreover, since in the limit $n\to \infty $ the linear span of
$\mathcal{S}_n$ coincides with the entire algebra of operators
generated by $\tilde{\mathcal{E}}$, this hierarchy embraces,
according to Lemma~\ref{lemeq}, all the conditions that can be
built from our approach. The strategy that we propose to verify
the quantum origin of a given behavior $P$ is thus the following.
Check first if there exists a certificate $\Gamma^1$ of order $1$
associated to $P$. If there is no such certificate, we can
conclude that the behavior $P$ is not quantum, otherwise check the
existence of a certificate $\Gamma^2$ of order 2. Repeat the
procedure with certificates of increasing order as long as the
behavior $P$ satisfies the previous tests.

\begin{center}
\begin{figure}
\centering
  \includegraphics[width=6 cm]{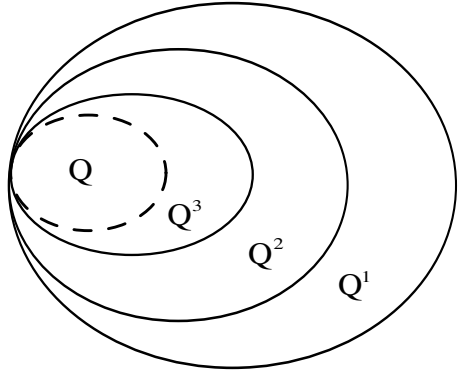}
  \caption{Geometrical interpretation of our hierarchy.
  $Q$ is the set of quantum behaviors. $Q^n$ denotes the set of all
  behaviors for which there exists a certificate of order $n$. Testing
 the existence of a certificate of order $n$ amounts to determine if a
 given behavior $P$ belongs to $Q^n$. Certificates of higher order
  provide a more accurate approximation of the quantum set $Q$,
  but are more demanding from a computational point of view.}
  \label{figgeom}
\end{figure}
\end{center}

A geometrical interpretation of our hierarchy is given in
Figure~\ref{figgeom}, where $Q^n$ denotes the set of all behaviors
$P$ for which there exists a certificate of order $n$.

\subsection{Sufficiency of the hierarchy}\label{suff}
We now show that our hierarchy is complete in the sense that
$\lim_{n\to\infty}Q^{n}=Q$, or in other words, that any
non-quantum behavior $P$ necessarily fails one of our conditions
at some step in the hierarchy.

\begin{theo}\label{theo}
Let $P$ be a behavior such that there exists a certificate
$\Gamma^n$ of order $n$ for all $n\geq 1$. Then $P$ belongs to
$Q$.
\end{theo}
\begin{proof}
The proof proceeds in two steps. We first show that the sequence
of certificates $\Gamma^n$ admits a proper limit
$\lim_{n\to\infty}\Gamma^n\to\Gamma^\infty$. We then construct
from the matrix $\Gamma^\infty$ a quantum state and quantum
operators acting on a (possibly infinite-dimensional) Hilbert
space $\mathcal{H}$ that reproduce the behavior $P$.

Note first, as shown in Appendix~B, that all the entries
$\{\Gamma^n_{s,t}\,:\,|s|,|t|\leq n\}$ of the matrices $\Gamma^n$
are bounded by 1, i.e., $|\Gamma^n_{s,t}|\leq 1$. Now, complete
each matrix $\Gamma^n$ with zeros to make it an infinite matrix
$\hat\Gamma^n$ with entries
$\{\hat\Gamma^n_{s,t}\,:\,|s|,|t|=0,1,\ldots\}$; we can then view
the matrices $\hat\Gamma^n$ as infinite vectors in $l_\infty$ (the
normed space of all bounded sequences $u=(u_1,u_2,...)$, with norm
given by $\|u\|_\infty=\sup_i|u_i|$). As the sequence
$\{\hat\Gamma^n \,:\, n=1,2\ldots\}$ belongs to the unit ball of
$l_\infty$, it admits, by the Banach-Alaoglu theorem, a
subsequence $\{n_i\}$ that converges in the weak-$\ast$ topology
to a limit $\hat\Gamma^{n_i}\to\Gamma^\infty$ when $i\to\infty$
\cite{reedsimon}. This implies in particular pointwise
convergence, i.e., \be\label{pwconv}
\lim_{i\to\infty}\hat\Gamma^{n_i}_{s,t}\to \Gamma^{\infty}_{s,t},
\ee
for all $s,t$. From the pointwise convergence, we deduce that
$\Gamma^\infty$ satisfies Eqs.~\eqref{certprob} and \eqref{ordern}
for all $s,t,u,$ and $v$. Moreover, let $\hat\Gamma^n_N$ denote
the submatrix of $\hat\Gamma^n$ corresponding to the entries
$\{\hat\Gamma^n_{s,t}\,:\,|s|,|t|\leq N\}$. Since
$\hat\Gamma^n_N\succeq 0$ for all $n$ and $N$, \eqref{pwconv}
implies that $\Gamma^\infty_N\succeq 0$ for all $N=1,2,\ldots$

In the remaining of the proof, we construct from the matrix
$\Gamma^\infty$ a state $\ket{\phi}$ and operators $\{\hat
E_a\,:\,a\in \tilde A\}$ and $\{\hat E_b\,:\,b\in \tilde B\}$
satisfying the properties of Definition~2.

The fact that $\Gamma^\infty_N\succeq 0$ for all $N$ implies that
there exists an infinite family of vectors $\{\ket{v_s}\,:\,
|s|=0,1,2,\ldots\}$ whose scalar products reproduce the entries of
$\Gamma^\infty$, i.e., \be\label{prodv}
\Gamma^\infty_{s,t}=\braket{v_s}{v_t} \ee for any $s,t$ of length
$|s|,|t|=0,1,2\ldots$ One way to establish this fact is through a
sequential Cholesky decomposition of the matrices
$\Gamma^\infty_N$~\cite{manal}.

We now take as our Hilbert space $\mathcal{H}$ the vector space
spanned by the vectors $\ket{v_s}$ and define, for all $a\in
\tilde A$, projectors $\hat E_a$ as follows \be \hat
E_a=\text{proj}\left(\text{span}\left\{\ket{v_{as}}\,:\,|as|=1,2\ldots\right\}\right)
\ee where $\text{proj}(V)$ is the projector on the subspace $V$.
Since $(E_a S)^\dagger E_{a'} T=\delta_{a,a'}S^\dagger E_a T$ when
$X(a)=X(a')$, it follows from \eqref{ordern} and \eqref{prodv}
that $\langle v_{as}|v_{a't}\rangle=\delta_{a,a'}\langle
v_s|v_{at}\rangle$, which in turn implies that \be \hat
E_a\ket{v_{a's}}=\delta_{aa'}\ket{v_{as}} \quad \text{if
}X(a)=X(a') \ee An immediate consequence of this is that
\be\label{faort} \hat E_a\hat E_{a'}=\delta_{aa'}\hat E_{a} \quad
\text{if }X(a)=X(a') \ee i.e, that the operators $\{\hat E_a\,:\,
a\in \tilde X\}$ form an orthogonal set of projectors. They thus
satisfy properties \emph{1} and \emph{2} of Definition~2.

Let us now examine the action of $\hat E_a$ over an arbitrary
vector $\ket{v_s}$. We find that
\begin{alignat}{2}
\hat E_a\ket{v_s}&=\hat E_a\ket{v_{as}}+\hat E_a\left(\ket{v_s}-\ket{v_{as}}\right)\nonumber\\
&=\ket{v_{as}}+\hat E_a\left(\ket{v_s}-\ket{v_{as}}\right)\nonumber\\
&=\ket{v_{as}}\label{fa}
\end{alignat}
The last identity follows from the fact that $\langle
v_{at}|v_s\rangle-\langle v_{at}|v_{as}\rangle=0$ which can be
deduced from \eqref{ordern}, \eqref{prodv}, and the relation
$\left(E_aT\right)^\dagger S-\left(E_aT\right)^\dagger E_aS=0$.
Property \eqref{fa} implies in particular that \be \hat
E_a\ket{v_1}=\ket{v_a}\label{fa1} \ee
By repeating the above
construction, we can build operators $\{\hat E_b\,:\, b\in\tilde
B\}$ for Bob that satisfy properties analogous to \eqref{faort},
\eqref{fa}, and \eqref{fa1}. From \eqref{fa}, \eqref{fa1}, and the
corresponding relations for Bob, we deduce by induction that
\be\label{sv} \hat S\ket{v_1}=\ket{v_{s}} \ee for any sequence
$\hat S$ of length $|\hat S|=0,1,2\ldots$ of the projectors
$\{\hat E_a\,:\, a\in\tilde A\}$ and $\{\hat E_b\,:\, b\in \tilde
B\}$. Combining Eqs.~\eqref{prodv} and \eqref{sv}, we find that
\begin{equation}
\Gamma^\infty_{s,t}=\bra{\phi}\hat S^\dagger \hat
T\ket{\phi}\label{sigmita}
\end{equation}
where we have defined $\ket{\phi}=\ket{v_1}$. Note that
$\ket{\phi}$ is a normalized vector since
$\braket{\phi}{\phi}=\Gamma^\infty_{1,1}=1$. Eqs.~\eqref{sigmita}
together with \eqref{certprob} imply that the state $\ket{\phi}$
and the operators $\hat E_a$ and $\hat E_b$ satisfy
Eqs.~\eqref{funda2}.

It now remains to verify property~3, i.e., that $[\hat E_a,\hat
E_b]=0$. From the relation $(E_{a} S)^\dagger E_{b} T-(E_{b}
S)^\dagger E_{a} T=0$, the properties~\eqref{ordern} satisfied by
$\Gamma^\infty$ and \eqref{sigmita}, we deduce that \be
\bra{\phi}\hat{S}^\dagger[\hat{E}_{a},\hat{E}_{b}]\hat{T}\ket{\phi}=0\label{commut}
\ee for any sequences $\hat S,\hat T$ of length $|\hat S|,|\hat
T|=0,1,2\ldots$ As the vectors $\hat S\ket{\phi}$ and $\hat
T\ket{\phi}$ span the support of the operators $\hat{E}_{a}$ and
$\hat{E}_{b}$, \eqref{commut} implies that the commutator $[\hat
E_a,\hat E_b]$ is equal to zero.
\end{proof}

\begin{cor}
$Q$ is a closed set.
\end{cor}
\begin{proof}
From Theorem~\ref{theo}, we know that $Q=\bigcap_{i=1}^\infty
Q^{i}$. As each of the sets $Q^{i}$ is closed, its infinite
intersection must be a closed set as well.
\end{proof}

\subsection{Stopping criteria and extraction of quantum state and measurements} \label{optimality}

Our hierarchy of conditions characterizes the quantum set $Q$ in an
asymptotic limit. Testing only a finite number of our conditions
may at most allow us to conclude that a given behavior \emph{does
not belong} to $Q$ (more precisely, testing the conditions up to
the $n^{th}$ step in the hierarchy allows us to detect all
behaviors that do not belong to $Q^n$). We now show that in
certain cases, it is possible to conclude at a finite order $n$ in
the hierarchy that a given behavior $P$ \emph{does belong} to $Q$
. In this case, we can also recover from the certificate
$\Gamma^n$ the quantum state $\ket{\psi}$ and the measurements
$E_a$ and $E_b$ reproducing the behavior $P$.

Let $\Gamma^n$ be certificate of order $n$ associated to the
behavior $P$. Fix a pair of inputs $X,Y$ and consider the set of
all sequences of the form $E_aE_bS$ where $a\in \tilde X$ and
$b\in \tilde Y$, together with all sequences of length $n-1$. Let
$J_{X,Y}$ be the set of indices associated to such sequences.
Define $\Gamma^n_{X,Y}$ as the submatrix of $\Gamma^n$ with
entries $\{\Gamma^n_{s,t}\,:\,s,t\in J_{X,Y}\}$. If
\begin{equation}
\mbox{rank}(\Gamma^n_{X,Y})=\mbox{rank}(\Gamma^n)\label{loop},
\end{equation}
for all $X,Y$, then we will say that the certificate $\Gamma^n$
has a \emph{rank loop}.

\begin{theo}\label{stop}
A behavior $P$ has a quantum representation of finite dimension if
and only if $P$ admits, for some finite $N$, a certificate
$\Gamma^N$ of order $N$ with a rank loop and
$\emph{rank}(\Gamma^N)\leq d$.
\end{theo}

Here, by a representation of dimension $d$, we mean that there exist
a quantum state $\ket{\Psi}\in{\cal H}$ and a set of operators
$\{E_a,E_b\in B({\cal H})\}$ satisfying the conditions of
Definition~\ref{defcomm2} for some Hilbert space $\mathcal{H}$ of
finite dimension $\text{dim}(\mathcal{H})=d$. We denote by $Q_d$ the
set of all behaviors having a $d$-dimensional quantum
representation.

\begin{proof}
We first prove the if $P$ has a finite dimensional representation,
there exists a certificate of order $n$ with a rank loop. As $P\in
Q_d$, there exist a state $\ket{\phi}\in{\cal H}$ and projective
measurements $E_\mu\in B({\cal H})$, as in Definition~2, for some
Hilbert space of $\text{dim}({\cal H})=d$. The matrix $\Gamma^n$
with entries $\Gamma_{s,t}^n=\bra{\phi}S^\dagger T\ket{\phi}$ for
all $S,T\in {\cal S}_n$ is clearly a certificate of order $n$
associated to $P$. Because ${\cal S}_n\subseteq {\cal S}_{n+1}$,
$\Gamma^n$ is a submatrix of $\Gamma^{n+1}$ for any $n$, and thus
$\text{rank}(\Gamma^n)\leq\text{rank}(\Gamma^{n+1}$). On the other
hand, the space generated by the vectors $S\ket{\phi}$, $S$ being
an arbitrary sequence has a dimension less or equal than
$\text{dim}(\mathcal{H})=d$ and therefore rank$(\Gamma^n)\leq d$
for all $n$. These two conditions imply that there exists an $N$
such that $\mbox{rank}(\Gamma^N)=\mbox{rank}(\Gamma^{N+1})\leq d$.
It follows that
$\mbox{rank}(\Gamma^{N+1}_{X,Y})=\mbox{rank}(\Gamma^{N+1}),$ for
all $X,Y$, and thus that $\Gamma^{N+1}$ has a rank
loop.\footnote{What we have proven is that $P$ has a, in general,
\emph{complex} rank looped certificate. To see that $P$ also has a
\emph{real} rank looped certificate, note that, for any set $n$,
$\mbox{Re}(\Gamma^n)$ is also a valid certificate for $P$. On the
other hand, rank($\mbox{Re}(\Gamma^n)$)$\leq 2\cdot
\mbox{dim}({\cal H})$, so the previous arguments can be applied to
$\mbox{Re}(\Gamma^n)$.}.

Let us now prove the converse statement. Suppose thus that $P$
admits a certificate $\Gamma^N$ with a rank loop and satisfying
$\text{rank}(\Gamma^N)=d$. Similarly to Section \ref{suff}, we can
perform a Cholesky decomposition of $\Gamma^N$ to write
$\Gamma^N_{st}=\braket{v_s}{v_t}$ for some finite set of vectors
$\{\ket{v_s}\,:\,|s|\leq N\}$, whose span is a vector space of
dimension at most $d$. Again as in Section \ref{suff}, we can then
define a set of operators $\hat{A}=\{\hat{E}_a\,:\,a\in\tilde{A}\}$
as
\be \hat
E_a=\mbox{proj}\left(\mbox{span}\left\{\ket{v_{as}}\,:\,|as|\leq
N\right\}\right). \ee It is easy to see that these projector
operators satisfy (\ref{faort}), and using the same arguments as
in section \ref{suff}, one can see that they also fulfill \be
\hat{E}_a\ket{v_s}=\ket{v_{as}}, \ee for $|as|\leq N$. In an
analogous way, we build operators
$\hat{B}=\{\hat{E}_b\,:\,b\in\tilde{B}\}$ for Bob with the same
properties. It is then immediate that
$\bra{v_1}\hat{S}^\dagger\hat{T}\ket{v_1}=\Gamma^N_{st}$, for
sequences $|\hat{S}|,|\hat{T}|\leq N$. In particular,
$\bra{v_1}\hat{E}_a\hat{E}_b\ket{v_1}=P(a,b)$. The operators in
$\hat{A}$ and $\hat{B}$ thus satisfy Eq.~\eqref{funda2} and
conditions \emph{1} and \emph{2} of Definition \ref{defcomm2}. It
remains to show that they also satisfy condition \emph{3}, i.e.,
commutativity.

Take any quadruple $a,b,X,Y$ such that $a\in \tilde{X},b\in
\tilde{Y}$, and consider the set of sequences ${\cal
S}_{XY}=\{S:|S|\leq N-1\mbox{ or } S=E_aE_bS',\mbox{ with } a\in
\tilde{X},b  \in \tilde{Y}, |S|\leq N\}$. Then, for any pair of
sequences $S,T\in {\cal S}_{XY}$,
\begin{equation}
\bra{v_s}\hat E_a\hat E_b-\hat E_b\hat
E_a\ket{v_t}=\Gamma^N_{as,bt}-\Gamma^N_{bs,at}=0,\label{commu}
\end{equation}
where the last equality comes from the constraints imposed on the
certificate $\Gamma^N$ by the operator identity $S^\dagger
E_aE_bT-S^\dagger E_bE_aT=0$. On the other hand, from condition
(\ref{loop}), we have that
\be \mbox{span}(\{\ket{v_s}:|s|\leq
N\})=\mbox{span}(\{\ket{v_s}:S\in{\cal S}_{XY}\}). \ee Since the
first set of vectors spans the support of the operators
$\hat{E}_a,\hat{E}_b$, relation (\ref{commu}) implies that
$\hat{E}_a,\hat{E}_b$ commute. As this holds for any quadruple
$a,b,X,Y$, it follows that $P\in Q_d$.
\end{proof}

\begin{cor}\label{cor}
Let $P$ be a behavior corresponding to a bipartite system where
Alice's (Bob's) measurements have $d_A$ $(d_B)$ possible outcomes
and such that each of the probabilities satisfies $P(a,b)>0$. Let
$\Gamma^2$ be a certificate of order 2 compatible with this
behavior. Then, rank$(\Gamma^2)=d_Ad_B$ implies that $P\in Q_d$,
with $d=d_Ad_B$.
\end{cor}

\begin{proof}
If $P(a,b)>0,\forall a\in A,\forall b\in B$, then, for any pair of
measurements $X,Y$, the $d_Ad_B$ vectors
$\{\ket{v_s}:s=\openone,E_a,E_b,E_aE_b:{a\in \tilde{X}, b \in
\tilde{Y}}\}$ can be shown to be linearly independent. This,
together with the fact that the rank of the whole matrix is equal to
$d_Ad_B$, implies that $\Gamma^2$ has a rank loop.
\end{proof}

The above theorem says that if our SDP outputs a certificate
$\Gamma$ with a rank loop, we know that $P$ belongs to $Q_d$, with
$d=\text{rank}(\Gamma)$. Moreover, from the proof of Theorem
\ref{stop} it is not difficult to see that we can even reconstruct
the state $\ket{\psi}$ and measurements $E_a$ and $E_b$ that yield
this finite-dimensional representation.

Given a behavior $P$ admitting a quantum representation of dimension
$d$, there may be, however, different certificates of order $n$
compatible with $P$, including some without rank loops. We have no
guarantee that our SDP will output a certificate that has a rank
loop, and thus in general we cannot guarantee that our hierachy of
SDP tests will stop after a finite number of iterations.

In view of this, it would be useful to incorporate some rank
minimization techniques in the implementation of our hierarchy.
That is, when checking the existence of certificates of order $n$,
we would like as well to minimize the rank of the corresponding
matrices. Indeed, let $\hat{\Gamma}^n$ be the certificate of order
$n$ for $P$ with minimum rank, and consider the series
$\hat{\Gamma}^1,\hat{\Gamma}^2,\hat{\Gamma}^3,...$ If
$\mbox{rank}(\hat{\Gamma}^{n+1})\not=\mbox{rank}(\hat{\Gamma}^n)$,
then
$\mbox{rank}(\hat{\Gamma}^{n+1})\geq\mbox{rank}(\hat{\Gamma}^n)+1$.
This, together with the fact that $\mbox{rank}(\hat{\Gamma}^n)\leq
d$ for all $n$, implies that there exists some $N\leq d$ such that
$\mbox{rank}(\hat{\Gamma}^{N+1})=\mbox{rank}(\hat{\Gamma}^N)$. On
the other hand, for all $X,Y$
\begin{equation}
\mbox{rank}(\hat{\Gamma}^N)\leq
\mbox{rank}(\hat{\Gamma}^{N+1}_{X,Y})\leq
\mbox{rank}(\hat{\Gamma}^{N+1}),
\end{equation}
and so
$\mbox{rank}(\hat{\Gamma}^{N+1})=\mbox{rank}(\hat{\Gamma}^{N+1}_{X,Y})$,
i.e., $\hat{\Gamma}^{N+1}$ has a rank loop.

Unfortunately, there are no known efficient methods to solve rank
minimization of positive semidefinite matrices with linear
constraints. There are, however, heuristics \cite{maryam} that
typically arrive at the optimal solution in just a few iterations.

\section{Applications}
\label{applications}

In this section, we present several applications of our method. We
first derive simple analytic conditions that are satisfied by all
quantum probabilities involving two measurements with two possible
outcomes. We then show how to apply our method to establish upper
bounds on the quantum violation of Bell inequalities.

From a general perspective, the hierarchy of necessary conditions
that we have introduced represents a systematic way of getting
better and better approximations to the set of quantum correlations.
Moreover, these approximations are nicely characterized in terms of
semidefinite constraints. Our method can thus be useful in any kind
of optimization problem over this set. This is particularly true
when we want to optimize the violation of Bell inequalities since
they are linear functions of the behaviors and thus the entire
optimization problem can be cast as a SDP.

Although the applications that we present here are restricted to a
bipartite scenario, our method can also be applied to a multipartite
scenario, e.g. see \cite{TV}.

\subsection{Analytic conditions for quantum behaviors with two inputs and two outputs}

Consider the measurement scenario described in Example 1 of Section
\ref{basicidea}, involving two measurements with two possible
outcomes for each observer. As we showed, a necessary condition for
a behavior to be quantum in this scenario is the existence of a
positive semidefinite matrix of the form (\ref{gammae1}). This
condition corresponds to the first one in our hierarchy and thus
characterize the set of behaviors $Q^1$. In the following, we
provide an analytic characterization of this set. The conditions
that we obtain can be interpreted as the quantum analogues of Bell
inequalities.

We make use of the following two lemmas:
\begin{lemma}\label{schurlemma}\emph{(Schur's lemma)}\cite{manal}
Let $M$ be a matrix such that
\begin{equation}
M=\left(\begin{array}{cc}P&Q\\Q^T&R\end{array}\right)\succeq 0,
\end{equation}
with $P\succ 0$. Then, $M\succeq 0$ if and only if
$R-Q^TP^{-1}Q\succeq 0$.
\end{lemma}

\begin{lemma}\label{landado}
Let $M_{z,t}$ be a real symmetric matrix of the form
\begin{equation}
M_{z,t}=\left(\begin{array}{cccc}1&z&x_1&x_2\\&1&x_3&x_4\\&&1&t\\&&&1\end{array}\right),
\label{paramatriz}
\end{equation}
with $|x_i|\leq 1$, $i=1,2,3,4$. Let
$f(x_1,x_2,x_3,x_4)=\arcsin(x_1)+\arcsin(x_2)+\arcsin(x_3)-\arcsin(x_4)$.
Then, there exists a pair of values $(z,t)$ such that
$M_{z,t}\succeq 0$ if and only if
\begin{equation}
|f(x_1,x_2,x_3,x_4)|\leq \pi \label{condiciones}
\end{equation}
for all possible permutations of $x_1,x_2,x_3,x_4$.
\end{lemma}
\begin{proof}
See Appendix~E.
\end{proof}

Now, apply Schur's lemma to matrix (\ref{gammae1}), taking the upper
block to be $P=1\succ 0$. It then follows that the positivity of
(\ref{gammae1}) is equivalent to the positivity of the matrix
$\Gamma'$ given by
\be\Gamma'=\begin{pmatrix}
1-C^2_1& u-C_1C_2& C_{13}-C_1C_3 & C_{14}-C_1C_4\\
& 1-C_2^2 & C_{23}-C_2C_3 & C_{24}-C_2C_4\\
& & 1-C_3^2 & v-C_3C_4 \\
& & & 1-C_4^2.
\end{pmatrix}
\ee
Note that we can restrict our analysis to the case where all the
elements in the diagonal are strictly positive. Indeed, if a
diagonal element is equal to zero, the corresponding measurement,
say by Alice, is deterministic, i.e., it always returns the same
outcome. Then, Alice is left with one effective measurement (at
most) and there always exists a classical, hence a quantum, model
for this type of scenario. Suppose thus that all the diagonal
elements of $\Gamma'$ are different from zero. Multiplying
$\Gamma'$ on both sides by the diagonal matrix
$M_{ii}=(1-C^2_i)^{-1/2}$ ($i=1,\ldots,4$), we obtain a matrix of
the same form as the one of Lemma (\ref{landado}). Applying this
lemma, we conclude, together with the previous observation, that a
necessary and sufficient condition for a behavior to belong to
$Q^1$ is either that there exist an $i$ such that $C_i^2=1$ or
that
\begin{equation}
\label{necond}
\left|\sum_{i,j}\arcsin\left(\frac{C_{ij}-C_iC_j}{\sqrt{(1-C_i^2)(1-C_j^2)}}\right)
-2\arcsin\left(\frac{C_{kl}-C_kC_l}{\sqrt{(1-C_k^2)(1-C_l^2)}}\right)\right|\leq
\pi
\end{equation}
for all $k=1,2$ and $l=3,4$. This condition is of course only a
necessary condition for quantum behaviors.

Note that a weaker necessary condition for a behavior to be quantum
follows from the positivity of
\be\Gamma=\begin{pmatrix}
& 1& u& C_{13} & C_{14}\\
& & 1 & C_{23} & C_{24}\\
& & & 1 & v \\
& & & & 1
\end{pmatrix} ,
\label{gammats} \ee which is simply a submatrix of \eqref{gammae1}.
A direct application of Lemma \ref{landado} implies that a behavior
is quantum if $|\sum_{ij}\arcsin(C_{ij})-2\arcsin(C_{kl})|\leq \pi$
for all $k=1,2$, $l=3,4$. This condition, which, as we said, is
weaker than \eqref{necond}, had previously been obtained in
\cite{tsir,Landau,Masanes2}.

\subsection{Quantum violation of Bell inequalities}\label{bi}

Bell inequalities are constraints satisfied by all behaviors that
originate from classical non-communicating observers. As mentioned
in the Introduction, for a finite number of measurements and
outcomes, the set of behaviors achievable using classically
correlated instructions (shared randomness) defines a polytope, that
is, a convex set with a finite number of extreme points (see also
Figure \ref{sets}). It can then alternatively be completely
characterized by a finite number of facets, which correspond to the
well-known Bell inequalities \cite{wernerwolf}. A given behavior $P$
thus admits a local classical model if and only if it satisfies all
the Bell inequalities. In the space of behaviors, a Bell inequality
can be viewed as a hyperplane that separates the space in two
regions. A generic Bell inequality can thus be written as
\begin{equation}\label{bellin}
   I(P)=\sum_{a,b}c_{ab}P(a,b)\leq I_C ,
\end{equation}
where $c_{ab}$ are the real coefficients defining the inequality
and $I_C$ is the maximal value achievable by local classical
points (and in particular which is attained by the extreme points
lying on the facet defined by the Bell inequality).

Since the work of Bell \cite{Bell}, we known that some quantum
behaviors are incompatible with a local classical description,
that is, that they violate a Bell inequality. This fact is often
referred to as quantum non-locality. In spite of many years of
work on quantum non-locality, there are no methods able to provide
the maximal quantum violation of a general Bell inequality, or
just non-trivial upper bounds to it \cite{lowerb}. An important
exception already mentioned in the introduction is the (tight)
bound derived by Tsirelson on the maximal violation of the CHSH
inequality.

Since our hierarchy of necessary conditions provides better and
better approximations to the set of quantum correlations, it can be
used to derive better and better upper bounds to the quantum
violation of a Bell inequality. Actually, our proof of completeness
guarantees the convergence to the maximal quantum value, that we
denote by $I_Q$. That is, by maximizing the value $I(P)$ of a Bell
inequality over the behaviors $P(a,b)\in Q^n$ admitting a
certificate of order $n$, one gets an upper bound $I_n$ to $I_Q$.
Clearly, we have that $I_1\geq I_2\dots \geq I_n\geq \dots\geq I_Q$
and $\lim_{n\to \infty}I_n\to I_Q$. Even while we are only able to
prove convergence to the quantum value in the asymptotic limit, the
quantum value or a very good upper-bound to it can often already be
obtained for a small relaxation order $n$, as we show in the
following.

The fact that Bell inequalities are linear functions of the joint
probabilities $P(a,b)$ signifies that we can cast the computation of
these upper bounds as SDP. Indeed, note that for any certificate
$\Gamma^n$, we can write the value $I(P)$ of a Bell inequality as
$I(P)=\tr(\beta_n\Gamma^n)$, where $\beta_n$ is a matrix whose
elements are all zero but the entries corresponding to
$\Gamma^n_{a,b}$. For instance, in the case $n=1$ one has (see
\eqref{exgamma}),
\begin{equation}\label{beta1}
    \beta_1=\frac{1}{2}\begin{pmatrix}
0 &C\\
C^T&0
\end{pmatrix} ,
\end{equation}
where $C$ is the matrix whose elements are the coefficients $c_{ab}$
in \eqref{bellin} defining the Bell inequality. Therefore the
calculation of $I_n$ amounts to solve the following SDP
\be
\begin{array}{ll}
\mbox{maximize}&\tr(\beta_n\Gamma^n)\\
\mbox{subject to}& \mbox{tr}\left(F_k^T\,\Gamma^n\right)=g_k(P) \quad k=1,...,m\label{existencebell}\\
& \Gamma^n\succeq 0
\end{array}\ee
In the remaining of this subsection we illustrate this approach by
applying it to several Bell inequalities.

But before presenting these results, let us make two technical
remarks. First, note that in the above optimization problems, we can
in general consider certificates that are intermediate between, say,
a certificate of order $1$ and a certificate of order $2$. Such a
certificate would be associated to a set of sequences of operators
$\mathcal{S}$ satisfying $\mathcal{S}_1\subset \mathcal{S}\subset
\mathcal{S}_2$. For instance, we could consider the set
$\mathcal{S}_{1+AB}=\mathcal{S}_1\cup \{E_aE_b ,:\,a\in \tilde A,
b\in \tilde B\}$ consisting of $\mathcal{S}_1$ together with all
products of one operator of Alice and one for Bob (while
$\mathcal{S}_2$ also contains product of two operators of Alice and
product of two operators of Bob). The corresponding bound $I_{1+AB}$
would then satisfy $I_1\leq I_{1+AB}\leq I_2$. In some cases this
bound might already be useful while requiring less computational
resources than $I_2$. In the following, we will therefore also
consider such bounds based on intermediate certificates. The
notation that we use is obvious, for instance $I_{1+AB+AA'B}$ is the
bound associated with the set of sequence operators
$\mathcal{S}_{1+AB+AA'B}=\mathcal{S}_{1}\cup
\{E_aE_b\}\cup\{E_aE_a'E_b\}$. Note that the rank loop conditions
derived in Subsection~4.2 generalize to the case of intermediate
certificates, see Appendix~C.

The second technical remark is that, as shown in Appendix~D, the
probabilities $P(a,b)=\Gamma^n_{a,b}$ corresponding to a certificate
$\Gamma^n$ are guaranteed to be positive only for certificates of
order $n\geq 2$ (or more generally for certificates associated with
set of operators $\mathcal{S}\supseteq\mathcal{S}_{1+AB}$). Thus,
when we maximize, as in \eqref{existencebell}, a Bell inequality
over all behaviors for which there exists a certificate $\Gamma^1$
of order 1, it may happens that the bound $I_1$ that we obtain
correspond to a solution with negative probabilities. By explicitly
adding to the SDP \eqref{existencebell}, the constraints
$\Gamma^1_{a,b}\geq 0$ that probabilities must be
positive\footnote{Adding such constraints leaves the optimization
problem in a SDP form}, we thus strengthen the upper bound $I_1$. In
the remaining of this section, when we mention an upper bound
obtained from a certificate of order 1, we always refer to this
strengthened version.

We start by analyzing the Collins-Gisin-Linden-Massar-Popescu
(CGLMP) family of Bell inequalities introduced in \cite{CGLMP}.
These inequalities are defined in a bipartite scenario where the
two observers can each make two measurements of $d$ outcomes. We
refer the reader to the original reference for the detailed
description of these inequalities. The inequality corresponding to
the case $d=2$ is the CHSH inequality. The best known lower bounds
on the quantum violation of these inequalities for $d\leq 8$ are
those given in Ref. \cite{ADGL}. The upper-bounds that we obtained
using our method are given in Table 1.
\begin{table}
\begin{center}\begin{tabular}{|c||c|c|c||c|c|c|}
\hline
  % after \\: \hline or \cline{col1-col2} \cline{col3-col4} ...
& \multicolumn{3}{|c||}{$I_1$} &\multicolumn{3}{c|}{$I_{1+AB}$}\\
\cline{2-7}
  $d$ & Value & Matrix Size & Rank Loop & Value & Matrix Size & Rank Loop\\
  \hline
  2 & 2.8284 & 5 & N/A& 2.8284 & 9 & Yes\\
  3 & 3.1547 & 9 & N/A& 2.9149 & 25 & Yes \\
  4 & 3.2126 & 13 & N/A& 2.9727 & 49 & Yes\\
  5 & 3.2997 & 17 & N/A& 3.0157 & 81 & Yes\\
  6 & 3.3378 & 21 & N/A& 3.0497 & 121 & Yes\\
  7 & 3.3843 & 25 & N/A& 3.0776 & 169 & Yes\\
  8 & 3.4115 & 29 & N/A& 3.1013 & 225 & Yes\\
    \hline
\end{tabular}
\caption{Upper bounds on the violation of the CGLMP inequality
derived from our construction. The local bound is equal to 2. The
upper bound $I_{1+AB}$ is already equal, up to numerical
precision, to the lower bounds given in \cite{ADGL}. We also
provide the size of the certificates in each case. Note that the
rank loop conditions defined in Subsection~\eqref{optimality} are
not applicable to certificates of order 1.}
\end{center}
\end{table}
Note first, that in the case $d=2$ (CHSH) the first certificate
already provides the actual quantum value, which is equal to the
Tsirelson bound. For $d$ larger than $2$, the quantum value is
recovered at the successive step corresponding to the certificate
$\Gamma^{1+AB}$. This can be seen by noting that the upper-bounds
$I_{1+AB}$ are equal to the lower bounds given in \cite{ADGL}.
Alternatively, one reaches the same conclusion by noting that the
stopping critera based on rank loops presented in Section
\ref{optimality} are satisfied. Thus, $\Gamma^{1+AB}$, and therefore
$\Gamma^2$, is already enough to get the maximal quantum violation
of CGLMP inequalities (at least until $d=8$) and certificates
$\Gamma^n$ with $n>2$ are redundant.

We have also considered other, perhaps less standard, Bell
inequalities, like the one presented in \cite{Pironio} (see also
\cite{us}) for the case in which Alice performs two measurements,
one of two outcomes and one of three outcomes, while Bob performs
three two-outcome measurements. The results are summarized in
Table~2. One can also get numerical lower bounds for the maximal
quantum violation for fixed dimension. In the case of qutrits, the
derived quantum violation is equal to 0.2532 \cite{Pironio,us}.
This is precisely the same value obtained when checking the last
certificate of Table~2. This certificate then, or equivalently
$\Gamma^3$, already provides a tight bound on the maximal quantum
violation. The same conclusion follows again by studying the rank
of the matrices appearing in these certificates.
\begin{table}
\begin{center}\begin{tabular}{|c|c|c|c|}
\hline
  % after \\: \hline or \cline{col1-col2} \cline{col3-col4} ...
  Upper bound & Value & Matrix Size & Rank Loop \\
  \hline
  $I_1$ & 0.3333 & 7 & N/A  \\
  $I_{1+AB}$ & 0.2653 & 16 & No  \\
  $I_{1+AB+AA'B}$ & 0.2532 & 22 & Yes  \\
  \hline
\end{tabular}
\label{tablestef} \caption{Upper bounds on the violation of the
$I_S$ inequality derived from our construction. The local bound is
equal to 0. The upper bound $I_{1+AB+AA'B}$ is already equal, up to
numerical precision, to the lower bound obtained numerically for
qutrits. We also provide the size of the certificates in each case.}
\end{center}
\end{table}

Finally, we also applied our techniques to the Froissard
inequality, also referred to as $I_{3322}$ inequality, given in
\cite{Froissard,CG}. Again, we refer the interested reader to
these references for the explicit form of the inequality. The best
known quantum violation of this inequality is equal to 0.25 in the
case of qubit systems, while the classical value is equal to zero.
By applying our hierarchy of conditions to this inequality, one
gets the upper bounds given in Table 3. Note that the values
derived for $\Gamma^2$ and $\Gamma^3$ are quite close and that no
rank loop is observed.\footnote{Rank loops should be considered in
a cautious way. Indeed it is sometimes difficult to numerically
distinguish a zero from a small eigenvalue.} It is remarkable that
none of our upper bounds coincides with the best known lower
bounds on the quantum violation, although they are very close to
it. This may be because in the case of this inequality our
hierarchy approaches more slowly the quantum solution, assuming it
to be equal to 0.25. However, one cannot exclude that the maximal
quantum violation of this inequality is obtained for systems of
dimension larger than two. Indeed, the existence of this type of
inequalities has recently been proven in \cite{PWPVJ,dimwitn,us}.
Thus, a quantum violation close to 0.2509 is perhaps attainable
beyond qubits.
\begin{table}\label{tablefro}
\begin{center}\begin{tabular}{|c|c|c|c|}
\hline
  % after \\: \hline or \cline{col1-col2} \cline{col3-col4} ...
  Upper bounds & Value & Matrix Size & Rank Loop \\
  \hline
  $I_1$ & 0.3333 & 7 & N/A  \\
  $I_{1+AB}$ & 0.2515 & 16 & No  \\
  $I_2$ & 0.25091 & 28 & No  \\
  $I_3$ & 0.25089 & 88 & No  \\
  \hline
\end{tabular}
\caption{Upper bounds on the violation of the $I_{3322}$ inequality
derived from our construction. The local bound is equal to 0.
Interestingly, none of our tests coincides with the best known lower
bound on the quantum violation, obtained for qubit systems. We also
provide the size of the certificates in each case.}
\end{center}
\end{table}

\section{Discussion and open questions}
\label{conclusion}

Characterizing the correlations attainable by quantum means is a
fundamental problem in Quantum Information Science and, more
generally, in Quantum Mechanics. To our knowledge, our construction
represents the only available tool to tackle this problem with full
generality: it applies to any number of parties, measurements and
outcomes. Moreover, the first steps in our hierarchy are easily
computable since they correspond to semidefinite programs of
reasonable size. Our construction provides a systematic way of
getting better and better approximations to the set of quantum
correlations and can be applied, for instance, to identify
correlations that do not admit a quantum representation, or to
estimate the maximum quantum violation of Bell inequalities.

In this work, after having presented in detail the hierarchy of
necessary conditions already introduced in \cite{anterior}, we
have (i) proven the completeness of the hierarchy, (ii) introduced
a criterion based on ranks loops that can guarantee at a finite
order in the hierarchy that a set of joint probabilities is
quantum, and we have shown in this case how to reconstruct the
quantum state and measurements reproducing these probabilities,
(iii) presented several examples illustrating the usefulness of
the method. Although our results are described in the bipartite
case, they can easily be extended to the multipartite scenario. To
conclude this work, we would like to go back to the commutation vs
tensor product issue briefly mentioned in
Subsection~\ref{defsets}, discuss the computational complexity of
our approach, and then present several open questions related to
the set of quantum correlations achievable with finite dimensional
Hilbert spaces.

We mention that it is possible to generalize the hierarchy
presented in this work and our proof of convergence to other
polynomial optimization problems with non-commutative variables
\cite{usprep}. A similar generalization was also recently
introduced in \cite{ito} to put upper-bounds on the entangled
value of quantum multi-player games. We also mention that an
alternative proof of convergence of our hierarchy is possible
using a result of Helton and McCullough \cite{helton}, as noted in
\cite{dohertyton,ito}.

\subsection{Commutation vs tensor product} \label{tensor}

There are two possible ways to impose that Alice and Bob perform
measurements on separated systems: through the condition that
their measurement operators commute, or through a tensor product
splitting of the whole Hilbert space. The two sets of quantum
correlations $Q$ and $Q'$ associated with each possibility are
defined in Subsection \ref{defsets}. Clearly, measurements that
have a product form commute with each other, and thus $Q'\subseteq
Q$. In the special case of finite-dimensional systems, one can in
fact show that both definitions are equivalent, i.e., $Q=Q'$
\cite{openproblem,prcomm} (see also \cite{vn}). For
infinite-dimensional systems whether they are equivalent or not
is, at the moment of writing, an open question \cite{openproblem}.
It is not even known if $Q'$ is dense in $Q$. (Note that the
statement in \cite{tsir} that the two sets are equivalent is
actually unproven \cite{prcomm}).

One can debate which definition should be regarded as the proper
one. Arguments in favor of the tensor product structure are
presented in \cite{openproblem}. Here, we have chosen commutativity
as this choice is consistent with the ethos adopted in this work.
Indeed, our main objective is to characterize the set of
correlations compatible with the general structure of Quantum Theory
but imposing as few additional constraints as possible: we impose no
restrictions on states, measurements, or even on the Hilbert space
dimension. In this spirit it should then be pointed out that there
exist in quantum field theories algebras of local (in the sense of
commuting) observables that cannot be split in a tensor product
structure\footnote{This result does not directly imply that $Q\neq
Q'$ since it could happen that all the correlations obtained by
performing commuting measurements on states belonging to these
spaces can also be realized in spaces with a tensor product
structure.}, yet in which it is possible to investigate the
correlations that can arise between two separated observers, and in
particular to study the amount by which Bell inequalities are
violated \cite{werner}. By investigating the structure of the set
$Q$ defined through commutativity, we are sure to include also these
examples and thus to deal with the most general correlations
compatible with Quantum Theory.

Of course, making the above distinction is only meaningful if $Q$
and $Q'$ happen to be distinct. But note that actually most of the
results of this work are independent of the definition chosen. As
$Q'\subseteq Q$, all the necessary conditions satisfied by points
in $Q$, in particular all the ones constituting the hierarchy, are
also valid for $Q'$. The stopping criteria presented in Subsection
\ref{optimality} are associated with correlations achievable with
finite-dimensional spaces, for which we known that $Q=Q'$, and
thus also apply to both cases. The unique distinction arises when
we consider the asymptotic behavior of our hierarchy: as our proof
of convergence to $Q$ explicitly use infinite-dimensional systems,
the hierarchy will also converge to $Q'$ only if $Q=Q'$ in the
most general setting. But for all practical applications of our
method where only a finite number of steps of the hierarchy are
involved, in particular for all numerical applications, one choice
of definition or the other does not make any difference. For
instance, all the results presented in Section \ref{applications}
apply equally well to both cases.

Note that it is not surprising that the limit of the hierarchy tends
to $Q$ rather than $Q'$, as the space separation between Alice and
Bob's measurements appears in the hierarchy only in the form of
constraints associated to the commutativity of these local
observables. For example, since $E_aE_bE_{a'}=E_aE_{a'}E_b$, we
impose that $\Gamma^n_{aba'}=\Gamma^n_{aa'b}$ for all $n$. If we
insist that the hierarchy should tend to $Q'$ rather than $Q$, it
will probably be necessary to add new constraints associated to the
tensor product structure. These constraints will have to reflect the
(at the moment unproven) differences, at the level of operator
algebra, between the commutation and tensor product case.

\subsection{Complexity of the hierarchy}
The computational complexity of our tests scales badly with the order
$n$ of the relaxation. For instance, in a measurement scenario with
$s$ inputs and $d$ outputs, it is not difficult to see that the size
of a certificate of order $n$ is roughly $(ds)^n$. The algorithms
used to solve the semidefinite programs associated with such
certificates have a running time that is polynomial in the size of
the matrix defining the SDP. Thus, using semidefinite programming to
decide if a certificate of order $n$ exists requires a time
exponential in $n$.

Note, however, that the numerical results presented in Section
\ref{bi} suggest that it might be sufficient, at least for some
families of measurement scenarios, to consider relaxations only up
to a bounded value $n$ to characterize, or obtain an already good
approximation, of the quantum set. Indeed, in the examples that we
considered, when maximizing the violation of Bell inequalities we
hit the quantum value, or obtained a very good upper-bound on it,
already at the second or third step in the hierarchy. The
suggestion that a finite number of steps of the hierarchy might
already characterize, or approximate well, the quantum region
turns out to be true in some particular case. For instance, for
measurement scenarios with two outputs, a result of Tsirelson
\cite{cir80} implies that deciding if a set of correlators (i.e.,
a quantity such as the $C_{ij}$ defined in Example 1 of Section 3)
is quantum can exactly be decided through semidefinite
programming, as noted by Wehner \cite{wehner}. The semidefinite
program considered by Wehner is a weaker version of the first step
of our hierarchy. In \cite{kempe2}, the authors show how for a
certain family of measurement scenario, corresponding to
\emph{unique games}, the quantum set can well be approximated
through semidefinite programming. The semidefinite programs
considered in \cite{kempe2} correspond again to the first step of
our hierarchy\footnote{With the additional constraint, when
maximizing the violation of Bell inequality, that the
probabilities must be positive, as mentioned in Section~5.2.}.

If all these results suggest that our construction might indeed
provide an efficient characterization of the quantum set for some
particular quantum scenarios, we do not expect this to be true in
full generality, as it has recently been shown, at least in the
tripartite case, that calculating the maximal quantum violation of a
Bell inequality is an NP-hard problem \cite{kempe}.

\subsection{Finite dimensional quantum systems}
In this work, we were mainly interested in characterizing the set
of quantum behaviors without any bound on the dimension of the
Hilbert space. We now present several open questions linked to the
finite-dimensional case.

\begin{itemize}
    \item Consider all possible quantum behaviors of $d$ outcomes where
the number of measurements is arbitrary. Gill recently asked
whether these correlations are attainable by measuring
$d$-dimensional quantum systems \cite{openproblem2}. The answer to
this question is no, as shown in \cite{PWPVJ} for the case of
three observers and in \cite{us,dimwitn} for bipartite systems.
Actually, no finite dimension is sufficient to generate the whole
set of quantum correlations of $d$ outcomes for three parties,
while the same result seems very plausible in the bipartite case
\cite{us,dimwitn}. Consider however a scenario where the number of
measurements is also finite. Are now all quantum correlations
(exactly) attainable by measuring a finite dimensional quantum
system?
\item Consider a measurement scenario with a finite number of inputs
and outputs. It is easy to see that in the tensor product scenario
discussed in Section~6.1. a quantum behavior can be approximated
arbitrarily well using finite-dimensional Hilbert spaces (see for
instance \cite{ito}). Does the same result hold in the commutative
case? If yes, then combining this result with the fact that $Q'=Q$
for finite-dimensional systems, would imply that $Q'$ is dense in
$Q$, and thus that our hierarchy converges to the quantum set $Q'$
defined through the tensor product structure.
\item What is the structure of the set of quantum behaviors corresponding to a Hilbert space
of fixed dimension $d$?  Very little is known in this case, we even
do not known if the corresponding quantum set is convex.
\item In relation with the above question, it would be interesting to understand how
    to incorporate in our hierarchy a bound on the Hilbert space dimension. It is in
    principle always possible to decide if a behavior can be
    represented with a Hilbert space of given dimension through semidefinite programming \cite{us} using known
    techniques of polynomial optimization \cite{lass,henr}. The
    corresponding SDPs, however, are very demanding from a
    computational point of view, much more than the one obtained
    from our construction where we do not bound the dimension.
    Can one modify our construction to design more efficient methods to approximate the set of correlations corresponding to $d$-dimensional
    quantum systems?

    As suggested by the results of Section 4.2,  a possibility would be to incorporate a bound on the rank of our
    certificates. There are, however, to our knowledge no efficient
    methods to solve SDPs with rank constraints. Is there any efficient way to relax these rank
    constraints to obtain good approximations to the set of quantum
    correlations with finite dimension?
\end{itemize}

\section{Acknowledgements}
We thank B. Tsirelson and R. Werner for useful discussions and
correspondence on the commutation vs tensor product question, and
T. Ito and B. Toner for pointing out to us the possibility of an
alternate proof of convergence of our hierarchy through the
results of Helton and McCullough. We acknowledge financial support
from the EU project QAP (IST-FET FP6-015848), from the Spanish
MEC, under FIS2004-05639, Consolider-Ingenio QOIT projects, and a
``Juan de la Cierva" grant, and from the Generalitat de Catalunya.

\appendix
\appendixpage
\renewcommand{\theequation}{A-\arabic{equation}}
\section{Basics of semidefinite programming}
\label{appsdp}

Semidefinite programming \cite{sdp} is a subfield of convex
optimization concerned with the following optimization problem,
known as the \emph{primal problem}
\begin{alignat}{2}
\mbox{maximize}\quad & \mbox{tr}(GZ)\nonumber\\
\mbox{subject to}\quad &\mbox{tr}F_iZ=c_i \quad i=1,...,p\label{dualb}\\
& Z\succeq 0\nonumber
\end{alignat}
The problem variable is the $n\times n$ matrix $Z$ and the problem
parameters are the $n\times n$ matrices $G,F_i$ and the scalars
$c_i$. A matrix $Z$ is said to be \emph{primal feasible} if it
satisfies the conditions expressed in (\ref{dualb}).

For each primal problem there is an associated \emph{dual
problem}, which is a minimization problem of the form
\begin{alignat}{2}
\mbox{minimize}\quad & c^Tx\nonumber\\
\mbox{subject to}\quad &F(x)=\sum_{i=1}^p x_iF_i-G\succeq 0
\label{primal}
\end{alignat}
where the variable is the vector $x$ with $p$ components $x_i$.
The dual problem is also a semidefinite program, i.e., it can be
put in the same form as \eqref{dualb}.  A vector $x$ is said to be
\emph{dual feasible} when $F(x)\geq 0$.

The key property of the dual program is that it yields bounds on the
optimal value of the primal program. To see this, take a dual
feasible point $x$ and a primal feasible point $Z$. Then $c^T
x-\mbox{tr}(GZ) = \sum_{i=1}^p\mbox{tr}(ZF_i)x_i - \mbox{tr}(GZ) =
\mbox{tr}(ZF(x))\geq 0$. This proves that the optimal primal value
$p^*$ and the optimal dual value $d^*$ satisfy $d^*\leq p^*$. In
fact, it usually happens that $d^*=p^*$. A sufficient condition for
this to hold is that there exists a strict feasible point of the
primal problem \cite{sdp}, that is, that there exists a matrix
$Z\succ 0$ that is primal feasible. Such a situation appears in the
SDP problem (\ref{existence}), as for any matrix $\Gamma$ satisfying
the corresponding linear constraints, we can always take $\lambda$
small enough so that $\Gamma-\lambda\openone\succ 0$.

There exist many available numerical packages to solve SDPs, for
instance for Matlab, the toolboxes SeDuMi \cite{sedumi} and YALMIP
\cite{yalmip}. Such algorithms usually solve both the primal and
the dual at the same time and thus yields bounds on the accuracy
of the obtained solution.

\section{Certificates have bounded entries}
\begin{prop}
Let $\Gamma^n$ be a certificate of order $n$ for a behavior $P$.
Then, $|\Gamma^n_{st}|\leq 1$, for all sequences $S,T$. That is,
the set of all certificates of order $n$ for $P$ is
bounded.\end{prop}

\begin{proof}
Because $\Gamma^n\succeq 0$, it just suffices to prove that all
diagonal elements are smaller or equal than 1. Consider thus any
$2\times 2$ submatrix of $\Gamma^n$:
\begin{equation}
\left(\begin{array}{cc}\Gamma^n_{ss}&\Gamma^n_{st}\\
\Gamma^n_{ts}&\Gamma^n_{tt}\end{array}\right).
\end{equation}
This submatrix must be positive semidefinite or, equivalently, its
coefficients have to satisfy $\Gamma^n_{ss},\Gamma^n_{tt}\geq 0$
and $\Gamma^n_{ss}\cdot \Gamma^n_{tt}\geq
\Gamma^n_{st}\cdot\Gamma^n_{ts}$. Now, take $T=E_a S$. From the
operator relation $S^\dagger T=T^\dagger S=T^\dagger T$, it
follows that $\Gamma^n_{st}=\Gamma^n_{ts}=\Gamma^n_{tt}$. This,
together with the positivity conditions, implies that
\begin{equation}
\Gamma^n_{tt}\leq \Gamma^n_{ss},\mbox{ for } T=E_a S, \forall
|S|\leq n,a\in\tilde{A}.
\end{equation}
In particular,
\begin{equation}
\Gamma^n_{aa}\leq \Gamma^n_{11}=1 \quad\text{for all } a.
\end{equation}
And, obviously, the same relations hold replacing $a$'s by $b$'s.
By induction, it is straightforward that $\Gamma^n_{ss}\leq
1,\forall S$, and, therefore, $|\Gamma^n_{st}|\leq 1,\forall S,T$.
\end{proof}

\section{Rank loop conditions for intermediate certificates}

We state here some results about rank loop conditions similar to the
ones introduced in Subsection~\ref{optimality} and which hold for
``intermediate certificates'' such as those that we used in
Subsection~\ref{bi} to maximize the violation of Bell inequalities.

Let us first define more precisely the certificates that we are
considering here. Given a pair of measurements $X,Y$, denote by
${\cal S}_{n+XY}$ the set of sequences ${\cal S}_n\cup \{S\in{\cal
S}_{n+1}\,:\,S=E_aE_bS',a\in \tilde{X},b\in \tilde{Y}\}$, i.e.,
$\mathcal{S}_{n+XY}$ is the set that contains all sequences of
length $n$ together with all the sequences of length $n+1$ that are
of the form $E_aE_bS'$ for some $a\in\tilde X,b\in\tilde Y$. It is
thus intermediate between the set of sequences of length $n$ and
$n+1$, as
$\mathcal{S}_n\subseteq\mathcal{S}_{n+XY}\subseteq\mathcal{S}_{n+1}$.
Given a vector $n$ of positive integers $n_{\textsc{xy}}$, define
${\cal S}_{n+AB}$ as the union of all sets ${\cal
S}_{n_\textsc{xy}+XY}$. By abuse of notation, when $n$ is an integer
we interpret it as the vector $(n,n,\ldots,n)$. With the notation
that we have just defined, we have for instance that ${\cal
S}_{1+AB}=\{\openone,E_a,E_b,E_a E_b:a,b\in \tilde{A},\tilde{B}\}$,
which is one of the set of sequences that we used in the numerical
applications presented in Subsection~\ref{bi}.

Given an arbitrary certificate $\Gamma$ associated to a set of
operators $\mathcal{S}$ and a vector $n$ of positive integers such
that $\mathcal{S}_{n+XY}\subseteq \mathcal{S}$, denote by
$\Gamma_{n+XY}$ the submatrix of $\Gamma$ corresponding to the set
of sequences $\mathcal{S}_{n+XY}$. Define similarly $\Gamma_{n+AB}$.
If there exists a vector $N$ such that
\begin{equation}
\mbox{rank}(\Gamma_{N_\textsc{xy}+XY})=\mbox{rank}(\Gamma_{N+AB})\label{loop2},
\end{equation}
for all $X,Y$, then we will say that the certificate $\Gamma$ has a
\emph{rank loop}. (Note that this definition is weaker than the one
given in Subsection~\ref{optimality}).

\begin{theo}\label{15}
A behavior $P$ has a quantum representation of finite dimension $d$
if and only if $P$ admits, for some $N$, a certificate $\Gamma$ with
a rank loop, and $\emph{rank}(\Gamma_{N+AB})\leq d$.
\end{theo}

\begin{cor}\label{16}
Let $P$ be a behavior corresponding to a bipartite system where
Alice's (Bob's) measurements have $d_A$ $(d_B)$ possible outcomes
and such that each of the probabilities satisfies $P(a,b)>0$, for
all $a\in A, b\in B$. Let $\Gamma$ be a certificate compatible with
this behavior associated to the set of operators ${\cal S}$, with
${\cal S}_{1+AB}\subseteq {\cal S}$. Then, rank$(\Gamma)=d_Ad_B$
implies that $P$ has a quantum representation of dimension $d_Ad_B$.
\end{cor}

The proofs of Theorem~\ref{15} and Corollary~\ref{16} follow along
the same lines as the proofs of Theorem~\ref{stop} and
Corollary~\ref{cor}.

\section{Certificates and non-negativity of probabilities}
\label{results}

Let $\mathcal{S}_{1+AB}=\{\openone,E_a,E_b,E_a E_b:a,b\in
\tilde{A},\tilde{B}\}$ be the set of all sequences of length less
than~$1$, together with all product operators consisting of one
operator of Alice and one of Bob. The proposition here below states
that the existence of a certificate $\Gamma$ corresponding to a set
of operators that contains $\mathcal{S}_{1+AB}$ as a subset, thus in
particular the existence of a certificate of order $n$ with $n\geq
2$, implies that the elements $P(a,b)$ of the behavior associated to
$\Gamma$ are proper probabilities, i.e., they are non-negative
numbers.

Before showing this, however, let us remind some notation introduced
in Subsection~\ref{defsets}. We defined a behavior as a set of joint
probabilities $P=\{P(a,b)\,:\,a\in A, b\in B\}$ and implicitly
assumed that they satisfy the no-signalling constraints
$P(a)=\sum_{b\in Y}P(a,b)$ and $P(b)=\sum_{a\in X}P(a,b)$. To remove
the redundancy associated with these constraints, we introduced the
reduced outcome sets $\tilde A$ and $\tilde B$ so that $P$ can be
alternatively represented as $P=\{P(a),P(b),P(a,b)\,:\,a\in\tilde A,
b\in \tilde B\}$. Having reminded this definition, it is now easy to
see that the sets of operators
$\mathcal{S}_{1+AB}=\{\openone,E_a,E_b,E_a E_b:a,b\in
\tilde{A},\tilde{B}\}$ and $\mathcal{S}_{AB}=\{E_aE_b\,:\,a\in A,
b\in B\}$ are linearly equivalent.

\begin{prop}\label{17}
Consider a measurement scenario $(A,B,\mathcal{X},\mathcal{Y})$, and
let $P=\{P(a,b):a\in A,b\in B\}$ be a set of real numbers. If there
exists a certificate $\Gamma$ for $P$ corresponding to a set
$\mathcal{S}$ such that $\mathcal{S}_{1+AB}\subseteq \mathcal{S}$,
then the numbers $P(a,b)$ represent proper probabilities, i.e.,
$P(a,b)\geq 0,$ for all $a$ and $b$.
\end{prop}
\begin{proof}
Let $P$ admit a certificate as in Proposition~\ref{17}. Then,
according to Lemma \ref{lemeq}, $P$ also admits a certificate
associated to the set ${\cal S}_{1+AB}$, and thus also a certificate
$\Gamma'$ associated to the set $\mathcal{S}_{AB}=\{E_aE_b\,:\, a\in
A,b\in B\}$. Since $\Gamma'\succeq 0$, its diagonal elements
$\Gamma'_{ab,ab}=P(a,b)$ must be non-negative.
\end{proof}

\section{Proof of Lemma \ref{landado}}\label{prooflandado}

\begin{proof}
Suppose that there exists a pair of values $(z,t)$, with $|t|<1$,
such that

\begin{equation}
M_{z,t}=\left(\begin{array}{cc}P&Q\\Q^T&R\end{array}\right)\succeq
0,
\end{equation}

\noindent with
$P=\left(\begin{array}{cc}1&z\\z&1\end{array}\right)$,
$Q=\left(\begin{array}{cc}x_1&x_2\\x_3&x_4\end{array}\right)$,
$R=\left(\begin{array}{cc}1&t\\t&1\end{array}\right)$. Because
$|t|<1$ implies $R\succ 0$, Lemma \ref{schurlemma} in section
\ref{applications} states that the positivity of $M_{z,t}$ is
equivalent to the condition $D\equiv P-Q^TQ^{-1}Q\succeq 0$. Now,
$D$ is a $2\times 2$ matrix with non diagonal free entries and so
its positive semidefiniteness is equivalent to demanding that
$D_{11},D_{22}\geq 0$. Therefore, we can get rid of the variable
$z$. Taking into account that $t^2-1< 0$, we have that both
conditions are equivalent to

\begin{eqnarray}
&\alpha_1\leq y\leq \alpha_2\nonumber\\
&\beta_1\leq y\leq \beta_2, \label{casa}
\end{eqnarray}

\noindent for
$\alpha_{1,2}=x_1x_2\mp\sqrt{x_1^2x_2^2-x_2^2-x_1^2+1}$,
$\beta_{1,2}=x_3x_4\mp\sqrt{x_3^2x_4^2-x_3^2-x_4^2+1}$.

It can be verified that $|\alpha_{1,2}|,|\beta_{1,2}|\leq 1$. A
solution $\max(\alpha_1,\beta_1)\leq t\leq
\mbox{min}(\alpha_2,\beta_2)$ can be found if and only if

\begin{eqnarray}
&\alpha_1\leq \beta_2\nonumber\\
&\beta_1\leq \alpha_2, \label{skinner}
\end{eqnarray}

\noindent and the requirement that $|t|<1$ translates into
$\max(\alpha_1,\beta_1),\mbox{min}(\alpha_2,\beta_2)$ are not both
equal to $\pm 1$.

Now, it can be proven that condition (\ref{skinner}) holds for
\emph{any} matrix $M_{z,t}$ for which there exists a pair of
values $z,t$ that makes it positive semidefinite. To see this,
notice that, for $M_{z,t}$ to be positive semidefinite it is
necessary that $|t|\leq 1$. So, if such a couple of values exist,
for any $\epsilon>0$ the matrix
$\frac{1}{\sqrt{1+\epsilon}}(M_{z,t}+\epsilon\openone)\frac{1}{\sqrt{1+\epsilon}}$
is of the form (\ref{paramatriz}) and there exists a pair of
values $(z'=z'/(1+\epsilon),t'=t/(1+\epsilon))$, with $|t'|<1$,
that make it positive semidefinite. Therefore, the vector
$(x_i/(1+\epsilon)$ has to satisfy (\ref{skinner}). Because this
holds for any $\epsilon>0$, by continuity, also the vector $(x_i)$
will satisfy (\ref{skinner}).

Next we will prove that any vector satisfying $(\ref{skinner})$
corresponds to a matrix of the type $M$ for which there exists a
couple of values $(z,t)$ that make it positive semidefinite.
Suppose, thus, that $(\ref{skinner})$ holds. Two situations can
arise: either
$\max(\alpha_1,\beta_1)=\mbox{min}(\alpha_2,\beta_2)=\pm1$ or not.
In the second case, we know that we can find a pair of values
$(z,t)$ such that $M_{z,t}\succeq 0$, whereas in the first case it
can be shown that $x_1=x_2=x;x_3=x_4=x'$. But a positive
semidefinite $M$ matrix for this case is given by the formula
$M=D\cdot\left(M^*+\mbox{diag}(\frac{1}{x^2}-1,
(\frac{1}{x'})^2-1,0,0)\right)\cdot D$, where
$D=\mbox{diag}(x,x',1,1)$ and $M^*\succeq 0$ is a $4\times 4$
matrix whose entries are all ones. Therefore, condition
(\ref{skinner}) is necessary and sufficient to guarantee the
existence of a pair of values $(z,t)$ such that $M_{z,t}\succeq
0$. Making the change of variables $x_i\to\sin(\phi_i)$ in
(\ref{skinner}) leaves us with (\ref{condiciones}).
\end{proof}

%\newpage
\bibliography{covariance2}

\begin{thebibliography}{48}
\bibitem{lowerb}
\bibinfo{note}{{L}ower bounds to the maximal quantum violation can easily be
  obtained by searching specific states and measurements that violate the
  inequality}.
\end{thebibliography}

\end{document}